\documentclass[amsmath,amssymb,pra,reprint,doublecolum,aps]{revtex4-2}

\usepackage{verbatim}
\usepackage{ulem}
\usepackage{cancel}
\usepackage{color}
\usepackage{graphicx}
\usepackage{physics}
\usepackage{dcolumn}
\usepackage{bm}

\begin{document}


\title{Daemonic quantum battery charged by thermalization}
\author{Matias Araya Satriani}
 \email{satriani@ug.uchile.cl}
\author{Felipe Barra}%
 \email{fbarra@dfi.uchile.cl}
\affiliation{Departamento de F\'{i}sica, Facultad de Ciencias F\'{i}sicas y Matem\'{a}ticas, Universidad de Chile, 837.0415 Santiago, Chile}

\date{\today}

\begin{abstract}

The reduced state of a small system strongly coupled to a charger in thermal equilibrium may be athermal and used as a small battery once disconnected. By harnessing the battery-charger correlations, the battery's extractable energy can increase above the ergotropy.
We introduce a protocol that uses a quantum system as a memory that measures the charger and leaves the battery intact in its charged state. Using the information gained from the measurement, the daemonic ergotropy of the battery is extracted. Then the battery is reconnected to the charger, thermalizing and charging it. However, the memory should return to its initial standard state to close the thermodynamic cycle. Thus, on the one hand, the work cost of the cycle is the sum of the disconnecting and reconnecting battery-charger work plus the measurement and erasure work. On the other hand, the extracted energy is the daemonic ergotropy of the battery plus the ergotropy of the memory. The ratio of these quantities defines the efficiency of the cycle. The protocol is exemplified by a modified transverse spin 1/2 Ising chain, one spin functioning as the battery and the others as the charger. The memory is another auxiliary spin 1/2. We found pairs of measurement schemes from which we extract the same daemonic ergotropy from the battery, they dissipate the same amount of energy, and one leaves the memory in an active state, the other in a passive state. We study the memory's ergotropy and the daemonic ergotropy of the battery. We find that with measurements, the efficiency can surpass that of the unmeasured protocol, given conditions on temperature, coupling, and choice of the measurement operators. 
\end{abstract}

\maketitle


\section{Introduction} \label{intro}

Motivated by the growing interest in quantum technologies, extensive research has been conducted on quantum machines over the past decade. Quantum batteries~\cite{Alicki2013, Binder2015, Barra2019, Farina2019, Pirmoradian2019, stableadiabatic, Hovh2020CabT, Zhao2021, Jing2021, Carrasco2022, Yao2022, Hu_2022, Barra22QBPHT, micromaser1, Morrone_2023, Yangyang2023, Mazzoncini2023, micromaser2}, which are storing energy systems, are an example. In this context, efforts have been directed at characterizing the amount of energy (charge) the battery can deliver, the speed at which this can be done, and proposing physical setups with particular advantages~\cite{stableadiabatic, darkstates, micromaser1, micromaser2}. However, evaluating the cost of producing and keeping the charge is essential to designing a quantum battery. References~\cite{Hovh2020CabT,Barra22QBPHT,Feliu2024} partially address the problem by evaluating the thermodynamic cost of producing resourceful states, considering that the Gibbs thermal state~\cite{Gibbs} with a reference temperature is a free resource. However, it is assumed that the control of the Hamiltonian of the systems is available to the agent. The thermodynamic cycle in~\cite{Barra22QBPHT,Feliu2024} starts with a battery strongly coupled to a charger in the thermal state. Due to the strong coupling, the reduced state of the battery is not thermal and may have charge. The amount of charge is measured with the ergotropy~\cite{Ergotropy}. The battery must be disconnected from the charger to extract its ergotropy. Thus, the protocol proceeds by disconnecting, extracting, and finally, reconnecting the battery to the charger to close the cycle. The thermalization process charges the battery.

Here, we modify the cycle, including an auxiliary quantum system, the memory~\cite{Piechocinska2000, Sagawa2008, Jacobs, Funo2013} that interacts with the charger, which is then projectively measured~\cite{Sagawa2008,Sagawa2009,Jacobs, Jacobs2,Funo2013, Daemonic, Auffeves2017, Manzano2018, POVM}. Its state stores the outcome of a generalized measurement on the charger. In this way, the battery's state remains intact. Conditioned on the measurement result, a unitary operation extracts the battery's ergotropy. Its average over the probability of the measurement outcomes gives the daemonic ergotropy~\cite{Daemonic}, which is always bigger or equal to the ergotropy of the battery's state. This gain stems from the battery-charger correlations caused by their coupling. However, to account for the efficiency of this new cycle, we must include the energy cost associated with the measurement and memory erasure~\cite{Landauer, Piechocinska2000, Landauerreview, Barkeshli2005, LandauerExperiment}. Moreover, the post-measurement state of the memory could be active, so its ergotropy (not demonic) can also be extracted.
The energetics of the cycle are studied and compared with the non-measured case in a simple model consisting of a spin 1/2 chain in which the first spin is the battery, and the others are the charger. An extra spin 1/2 particle plays the role of the memory. The choice of measurement is crucial for the process's efficiency. We propose an argument for choosing one that significantly enhances efficiency above the unmeasured protocol. As part of our analysis, we show that with all gains and works considered, the Kelvin-Plank statement of the second law is always fulfilled. Moreover, in close similarity to the work in~\cite{Sagawa2008, Jacobs, Funo2013,Piechocinska2000} we can pinpoint the role that information gain and Shannon entropy~\cite{Sagawa2008, Maruyama2009, Jacobs, Funo2013, ToIrev, Perarnau-Llobet2015, Fuel} have on the energetics of our cycle.

The rest of the article is organized as follows. Section~\ref{Protocol} is divided in subsection~\ref{sec2a}, where we review the working cycle for the battery proposed in~\cite{Barra22QBPHT} and in subsection~\ref{QMMem} where the new cycle, introducing measurements and extracting the battery's daemonic ergotropy and memory's ergotropy is described. Three subsections in~\ref{QMMem} describe a particular but important case and discuss the important role of the dissipated work. Then, in section~\ref{QMBattery}, we describe the model we study numerically in section~\ref{Results}. We summarize and present our conclusions in section~\ref{Conclusions}.

\section{Charging quantum batteries by thermalization: Protocol without and with measurement} 
\label{Protocol}

\subsection{Battery without measurement}
\label{sec2a}

We present the thermodynamic cycle considered in~\cite{Barra22QBPHT,Feliu2024}. The protocol uses a strongly coupled battery-charger system, in which the charger functions as a heat bath~\cite{Hovh2020CabT,Barra22QBPHT} at inverse temperature $\beta = 1/T$. The state of the total battery-charger system is initially the Gibbs equilibrium state $ \hat{\rho}_{\rm tot} = e^{-\beta \hat{H}_{\rm tot}} / Z_{\rm tot}$ where the Hamiltonian 
$\hat{H}_{\rm tot} = \hat{H}_{0} + \hat{V}^{b:c}_{\rm int}$
is the sum of $\hat{H}_{0}=\hat{H}_{b} + \hat{H}_{c}$, the Hamiltonian of the battery $\hat{H}_{b}$ plus the Hamiltonian of the charger $\hat{H}_{c}$, and the battery-charger interaction $\hat{V}_{\rm int}^{b:c}$.
        The cycle consists of the following four strokes:
        
    \begin{enumerate}

        \item The battery is disconnected from the charger in a quench~\footnote{A smooth disconnecting protocol was studied in~\cite{Feliu2024}.} so that the total state of the system remains unchanged.

        The energy cost of the quench reads:
        \begin{equation}                
        W_{d} = \text{Tr}[\hat{H}_{0} \hat{\rho}_{\rm tot}] -
        \text{Tr}[(\hat{H}_{0} + \hat{V}^{b:c}_{\rm int}) \hat{\rho}_{\rm tot}].         
        \end{equation}

        Tracing out the charger, we obtain the reduced state of the battery after the quench $\hat{\rho}_{b} = \text{Tr}_{c}[\hat{\rho}_{\rm tot}]$. The interaction $\hat{V}_{\rm int}^{b:c}$ is such that the reduced state of the battery $\hat{\rho}_{b}$ is active.

        \item A unitary operation $\hat{U}_{\mathcal{E}}$ extracts the ergotropy $\mathcal{E}$ of $\hat{\rho}_{b}$. The total system change is $\hat{\rho}_{\rm tot} \rightarrow (\hat{U}_{\mathcal{E}} \otimes I_{c}) \hat{\rho}_{\rm tot} (\hat{U}_{\mathcal{E}} \otimes I_{c})^{\dag}$, where the identity $I_{c}$ acts over the space of the charger. The ergotropy is~\cite{Ergotropy}:

        \begin{equation}
            \mathcal{E} = \text{Tr}[\hat{H}_{0} (\hat{\rho}_{\rm tot} - (\hat{U}_{\mathcal{E}} \otimes I_{c})\hat{\rho}_{\rm tot}(\hat{U}_{\mathcal{E}} \otimes I_{c})^{\dag})].
        \end{equation}

        \item The battery is reconnected to the charger in a quench, so the total state of the system remains unchanged. The energy cost of this step reads:

            \begin{eqnarray}
                 W_{r} &=&
                  \text{Tr}[(\hat{H}_{0}+ \hat{V}_{\rm int}^{b:c}) (\hat{U}_{\mathcal{E}} \otimes I_{c})\hat{\rho}_{\rm tot}(\hat{U}_{\mathcal{E}} \otimes I_{c})^{\dag}] \nonumber\\&& - 
                  \text{Tr}[\hat{H}_{0} (\hat{U}_{\mathcal{E}} \otimes I_{c})\hat{\rho}_{\rm tot}(\hat{U}_{\mathcal{E}} \otimes I_{c})^{\dag}].
            \end{eqnarray}            

        The sign convention denotes work performed on the composite system ($W_{d}$ and $W_{r}$) and work extracted from the system ($\mathcal{E}$).
        
        \item The cycle is closed by letting the system thermalize. The charger is in weak contact with a super bath~\cite{Barra22QBPHT} at temperature $T$ (with negligible energy cost), and the cycle is ready to start from 1.
        
    \end{enumerate}

    The efficiency of the cycle is characterized by:

    \begin{equation}
        \eta = \frac{\mathcal{E}}{W_{d} + W_{r}},
    \end{equation}
with $0 \leq \eta \leq 1$ as a consequence of the second law of thermodynamics~\cite{Hovh2020CabT}. 

Interestingly, for any choice of phases $\{ e^{i \theta_{k}} \}$, the operator
    \begin{equation}
        \hat{U}_{\mathcal{E}} = \sum_{k} e^{i \theta_{k}} \ket{E_{k}} \bra{p_{k}},
        \label{uergo}
    \end{equation}
extracts the maximal amount of energy from $\hat{\rho}_{b}$~\cite{Barra22QBPHT}.
    We arrange the eigenstates $ \{ \ket{E_{k}} \} $ of $\hat{H}_{b}$ such that the corresponding eigenvalues $\{ E_{k} \}$, are ordered in ascending order $E_{1} < E_{2} < \cdots$. Similarly, the eigenstates $\{ \ket{p_{k}} \}$ of $\hat{\rho}_{b}$ associated to the eigenvalues $\{ p_{k} \}$ are ordered in descending order $p_{1} \geq p_{2} \geq \cdots $, so that $\hat{\rho}_{b} = \sum_{k} p_{k} \ket{p_{k}} \bra{p_{k}}$. After the extraction of the ergotropy, the state $\hat{U}_{\mathcal{E}} \hat{\rho}_{b} \hat{U}_{\mathcal{E}}^{\dag} = \sum_{k} p_{k} \ket{E_{k}} \bra{E_{k}}$ is passive~\cite{Lenard}.

    If the battery-charger system is correlated after disconnecting, the reconnecting work $W_{r} \equiv W_{r}(\{ \theta_{k} \}) $ depends on the phases $\{ e^{i \theta_{k}} \}$. These phases can be taken so that $W_{r}$ is minimized and $\eta$ is maximized~\cite{Barra22QBPHT}.

\subsection{Daemonic quantum battery} \label{QMMem}

The protocol described above exploits correlations through the phases $\{\theta_k\}$ to minimize the reconnecting work. However, leveraging the correlations can increase the extractable energy above the ergotropy~\cite{Daemonic}. To do this, one relies on measurements~\cite{Jacobs, Funo2013, Fuel, Daemonic, Auffeves2017, Cottet2017, Manzano2018, ChargwithLFb, DaemonicCont, POVM, Yao2022, Yan2023}. So, in this sub-section, we include measurement and feedback in the energetics of the previous protocol. An auxiliary quantum system, the memory, interacts with the charger, building correlations between them. Following this interaction, a projective measurement is done on the memory.
Importantly, the measurement back-action affects the state of the charger but not the state of the battery, which we do not wish to disturb. The outcome of this measurement correlates with the state of the charger and, by extension, with the state of the battery. In this way, by accessing the states of the memory, one can acquire information about the state of the battery-charger system. We now present the working protocol that realizes the measurement.

    The Hamiltonian of the memory is $\hat{H}_{m} = \sum_{j} m_{j} \ket{m_{j}} \bra{m_{j}}$. It starts prepared in the initial standard state $\hat{\rho}_{m}^{0} = \ket{m_{0}} \bra{m_{0}}$, where $\ket{m_{0}}$ is the lowest energy eigenstate. The initial Hamiltonian of the whole system is:
    \begin{eqnarray}
        \hat{H}_{\rm tot} = \hat{H}_{b} + \hat{H}_{c} + \hat{V}^{b:c}_{\rm int} + \hat{H}_{m},
    \end{eqnarray}
with an initial state:
    
    \begin{equation}        
        \hat{\rho}_{\rm tot} = \frac{e^{-\beta (\hat{H}_{b} + \hat{H}_{c} + \hat{V}^{b:c}_{\rm int})}}{Z_{b:c}} \otimes \ket{m_{0}} \bra{m_{0}}.
    \end{equation}
 We now execute a generalized measurement on the charger. The measurement is realized by a unitary evolution $\hat{U}_{\rm int}^{c:m}$ that acts on the Hilbert spaces of the charger and the memory, followed by a projective measurement $P_{j} = \ket{m_{j}} \bra{m_{j}}$ done on the memory~\cite{Sagawa2009}. The operator $\hat{U}_{\rm int}^{c:m}$ correlates the charger and the memory and leaves the battery intact.
    
    The total state of the system after the measurement given the outcome $j$ is (from now on, we omit the tensor product with the identities of the missing Hilbert spaces for simplicity):

    \begin{eqnarray}
        \hat{\rho}_{\rm tot}^{(j)} &=& \frac{P_{j} \hat{U}_{\rm int}^{c:m} \hat{\rho}_{\rm tot} \hat{U}_{\rm int}^{c:m \dag} P_{j}}{\text{Tr}[P_{j} \hat{U}_{\rm int}^{c:m} \hat{\rho}_{\rm tot} \hat{U}_{\rm int}^{c:m \dag}]} \nonumber \\ &=& \frac{1}{{p_{j}}} M_{j} \hat{\rho}_{\rm b:c} M_{j}^{\dag} \otimes \ket{m_{j}}\bra{m_{j}},
    \end{eqnarray}
    where we have identified the measurement operators that act on the charger $M_{j} = \bra{m_{j}}\hat{U}_{\rm int}^{c:m}\ket{m_{0}}$ and the reduced battery-charger state $\hat{\rho}_{b:c} = \text{Tr}_{m}[\hat{\rho}_{tot}]$. 
    Finally, the probability of obtaining the outcome $j$ is $p_{j} = \text{Tr}[E_{j} \hat{\rho}_{c}]$, where $\hat{\rho}_c={\rm Tr}_b\hat{\rho}_{b:c}$ and $E_{j} = M_j^\dag M_j$.

    After the measurement, the battery is ready to keep the cycle going with the additional knowledge given by the measurement. We describe the energetics of the protocol. If the outcome registered on the memory is $j$, then the work cost of measurement is $\text{Tr}[\hat{H} \hat{\rho}_{\rm tot}^{(j)}] - \text{Tr}[\hat{H} \hat{\rho}_{\rm tot}]$ and in average
    \begin{equation}
        W_{\text{meas}} = \sum_{j} p_{j} (\text{Tr}[\hat{H} \hat{\rho}_{\rm tot}^{(j)}] - \text{Tr}[\hat{H} \hat{\rho}_{\rm tot}]),
    \end{equation}
    where $\hat{H} = \hat{H}_{\rm tot}$ if the measurement is made before the disconnection, and $\hat{H} = \hat{H}_{0} + \hat{H}_{m}$ if it is made after the disconnection.     

    The disconnecting work also depends on the order of events:

    \begin{eqnarray}\label{MeasDiss}
        W_{d} = -\text{Tr}[\hat{V}_{\rm int}^{b:c} \hat{\rho}],
        \label{wdis}
    \end{eqnarray}
    where $\hat{\rho} = \hat{\rho}_{b:c}$ if the disconnection is done before the measurement and $\hat{\rho} = \hat{\rho}_{b:c}^{(j)}=M_j \hat{\rho}_{b:c}M_{j}^{\dag}/p_j$ if it is done after the measurement. In this case, the average disconnecting work is given in \eqref{wdis} with $\hat{\rho} = \sum_j p_j\hat{\rho}_{b:c}^{(j)} = \sum_j M_j \hat{\rho}_{b:c}M_{j}^{\dag}$.

    Next, given the outcome $j$, the operator $\hat{U}_{{\mathcal E}_b}^{(j)}$ extracts the ergotropy $\mathcal{E}_{b}^{(j)}$ of the battery in the state $\hat{\rho}_{b}^{(j)} = \text{Tr}_{c,m}[\hat{\rho}_{\rm tot}^{(j)}]$, leaving the whole system in the state $\hat{U}_{{\mathcal E}_b}^{(j)} \hat{\rho}_{\rm tot}^{(j)}\hat{U}_{{\mathcal E}_b}^{(j)\dag} $. By averaging $\mathcal{E}_{b}^{(j)}$ over all possible outcomes we obtain the daemonic ergotropy~\cite{Daemonic}:
    \begin{equation}\label{DaemonicErg}
        \mathcal{E}_{b} = \sum_j p_j\mathcal{E}_b^{(j)} = \sum_{j} p_{j} \text{Tr}[\hat{H}_{b}(\hat{\rho}_{b}^{(j)} - \hat{U}_{{\mathcal E}_b}^{(j)} \hat{\rho}_{b}^{(j)}\hat{U}_{{\mathcal E}_b}^{(j)\dag})].
    \end{equation}

     The extra (always positive) energy extracted over the ergotropy due to the measurement is called \textit{daemonic gain} $\Delta \mathcal{E}_b = \mathcal{E}_b - \mathcal{E} \geq 0$~\cite{Daemonic}, where $\mathcal{E}$ is the ergotropy extracted without measurement.
    
    Now, we focus on the treatment of the memory after the measurement. From an observer with no information about the measurement, the total state of the system is a mixture of all possible outcomes. Thus, the reduced state of the memory would be $\hat{\rho}_{m}^{\prime} = \sum_{j} p_{j} \ket{m_{j}} \bra{m_{j}}$, which could be active. Consequently, its ergotropy could be extracted to achieve the highest possible process efficiency. We call this ergotropy $\mathcal{E}_{m}$. 
    The protocol for extracting the daemonic ergotropy from the battery uses the information stored in the memory. To extract the ergotropy daemonically from memory, an external agent would need to harness the information in it, thus creating the need for secondary memory. To keep the cycle closed (without the need for another system to store information), the ergotropy of the memory is extracted by an external agent with no information about the exact outcome, only the mixture the system is left in. Thus, the ergotropy of the memory, extracted by the operator $\hat{U}_{{\mathcal E}_m}$, is

    \begin{eqnarray}
        \mathcal{E}_{m} = \text{Tr}[\hat{H}_{m}(\hat{\rho}_{m}^{\prime} -  \hat{U}_{{\mathcal E}_m} \hat{\rho}_{m}^{\prime}\hat{U}_{{\mathcal E}_m}^{\dag})].
        \label{mem-ergo}
    \end{eqnarray}

    With some algebra, the total ergotropy extracted reads:

    \small
    \begin{equation}\label{totalergo}
        \mathcal{E}_{b} + \mathcal{E}_{m} = \sum_{j} p_{j} \text{Tr}[(\hat{H}_{0} + \hat{H}_{m}) (\hat{\rho}_{\rm tot}^{(j)} - \hat{U}_{{\mathcal E}_b}^{(j)} \hat{U}_{{\mathcal E}_m}\hat{\rho}_{\rm tot}^{(j)}\hat{U}_{{\mathcal E}_b}^{(j)\dag} \hat{U}_{{\mathcal E}_m}^{\dag})].
    \end{equation}
    \normalsize

The reconnecting work is drastically affected by the measurement. For example, if the measurement operators are rank-1 projectors, which is not the generic situation, then each outcome results in a product state between the charger and the battery, thus $\hat{U}_{\mathcal{E}_{b}}^{(j)}  \hat{\rho}_{b:c}^{(j)}\hat{U}_{\mathcal{E}_{b}}^{(j)\dag}$ and consequently, the reconnecting work $\text{Tr}[\hat{V}_{\rm int}^{b:c} \hat{U}_{\mathcal{E}_{b}}^{(j)}  \hat{\rho}_{b:c}^{(j)}\hat{U}_{\mathcal{E}_{b}}^{(j)\dag}]$
are independent of the phases $\{ \theta_{k} \}$ in \eqref{uergo}. On the other hand, if after the measurement, the state $\hat{\rho}_{b:c}^{(j)}$ is correlated, then the reconnecting work will depend on the chosen $\theta_{k}$, thus opening up the possibility for optimization~\cite{Barra22QBPHT}. In any case, the average reconnecting work is
  \begin{eqnarray}\label{wreco}
                 W_{r}(\{ \theta_{k} \}) & =& \sum_j p_j
                  \text{Tr}[\hat{V}_{\rm int}^{b:c} \hat{U}_{\mathcal{E}_{b}}^{(j)}  \hat{\rho}_{b:c}^{(j)}\hat{U}_{\mathcal{E}_{b}}^{(j)\dag}]\nonumber \\ &=& \sum_j p_j
                  \text{Tr}[\hat{U}_{\mathcal{E}_{b}}^{(j)\dag}\hat{V}_{\rm int}^{b:c} \hat{U}_{\mathcal{E}_{b}}^{(j)}  \hat{\rho}_{b:c}^{(j)}].
            \end{eqnarray}   
    
    The memory implementation has an additional work cost in the thermodynamic cycle. It must be reset to its initial state $\hat{\rho}_{m}^{0} = \ket{m_{0}} \bra{m_{0}}$, which is not in thermal equilibrium. It is also important to note that the passive state of the memory is generally not in thermal equilibrium with the heat bath of inverse temperature $\beta$. The minimal work needed to change the passive state of the memory to its initial state in the presence of a heat bath (e.g., the super bath) at temperature $T$ is determined by the difference in non-equilibrium free energy  $F(\hat{\rho}; \hat{H}) = \text{Tr}[\hat{H} \hat{\rho}] - TS(\rho),$ where $S(\rho)=-\text{Tr}[\rho\ln\rho]$ is the von Neumann entropy. As shown in~\cite{Esposito2011, ToIrev}, the work done on a system that changes from some state $\hat{\rho}^{\prime}$ with Hamiltonian $\hat{H}^{\prime}$ to another state $\hat{\rho}$ with Hamiltonian $\hat{H}$ in contact with an ideal heat bath at temperature $T$ satisfies:
    
    \begin{eqnarray}\label{Noneq2law}
        W \geq F(\hat{\rho}, \hat{H}) - F(\hat{\rho}^{\prime}, \hat{H}^{\prime})\equiv \Delta F.
    \end{eqnarray}

    Following this, the minimum work required in the reset process of the memory is given by the change $\Delta F_{\text{reset}}$ :

\begin{equation}\label{FreeEreset}
        W_{\rm reset}\geq\Delta F_{\text{reset}}  = F(\hat{\rho}_{m}^{0}, \hat{H}_{m}) - F(\hat{U}_{{\mathcal E}_m} \hat{\rho}_{m}^{\prime}\hat{U}_{{\mathcal E}_m}^{\dag}, \hat{H}_{m}).
\end{equation}    
    Since $S(\hat{\rho}_{m}^{0})=0$ and $S(\hat{U}_{{\mathcal E}_m} \hat{\rho}_{m}^{\prime}\hat{U}_{{\mathcal E}_m}^{\dag})=S(\hat{\rho}_{m}^{\prime})=H(X^{m})$ is given by the Shannon entropy $H(X^{m})=-\sum_{j}p_j \ln{p_j}$, the lower bound of the memory reset work~\cite{Sagawa2009, Esposito2011} can be written as:

    \begin{equation}
        W_{\text{reset}} \geq   T H(X^{m})+\text{Tr}[\hat{H}_{m}(\hat{\rho}_{m}^{0} - \hat{U}_{{\mathcal E}_m} \hat{\rho}_{m}^{\prime}\hat{U}_{{\mathcal E}_m}^{\dag})].
        \label{resetbound}
    \end{equation}
Note that the second term at the right-hand side of the inequality is negative if $\hat{\rho}_{m}^{0}$ is the state of minimum energy for the battery (as we consider) or vanishes if all energy states are degenerate.    

    After resetting the memory and allowing the battery and charger to thermalize, the cycle is closed and ready to start again.
    
    The total work done in the cycle is:

    \begin{equation}
        W_{\rm tot} = W_{d} + W_{r}(\{ \theta_{k} \}) + W_{\text{meas}} + W_{\text{reset}}.
    \end{equation}

    The total ergotropy extracted in the cycle is:

    \begin{equation}
        \mathcal{E}_{\rm tot} = \mathcal{E}_{b} + \mathcal{E}_{m}.
    \end{equation}

    So the efficiency is:

    \begin{equation}
        \eta_{\text{meas}} = \frac{\mathcal{E}_{\rm tot}}{W_{\rm tot}} \leq 1. \label{Etaquantum}
    \end{equation}
 The inequality follows from Eq.\eqref{WdissGeneral}, which is demonstrated in Appendix \ref{AB}.

\subsubsection{Particular case:  fully degenerate memory} \label{CLMem}

     We present a particular case for the memory system shown in the previous section, which we call degenerate memory. For this case, the Hamiltonian of the memory system is considered to be fully degenerate. Given this full degeneracy, no energy cost is associated with the changes in the memory's internal energy after the measurement. Therefore, it can not store ergotropy. The extraction of the battery's daemonic ergotropy (Eq. \eqref{DaemonicErg}), the disconnection work, and the connection work of the battery-charger system are independent of the structure of the memory. 
     
     The most important consequence of this simplification is that the bound for the resetting work~\cite{Landauer, Piechocinska2000} simplifies to Landauer's bound
    \begin{eqnarray}
        W_{\rm reset} \geq T H(X^{m}). 
    \end{eqnarray}

\subsubsection{Comments on the thermodynamics of measurement, feedback and memory reset}\label{Comment}
  Similarly to the lower bound of the reset work, a lower bound for the measurement work done in the memory system in contact with a heat bath has been derived in Ref.~\cite{Sagawa2009}. This bound would relate the non-equilibrium free energy change~\cite{ToIrev} of the memory during the measurement process and the information about the battery-charger system acquired by the measurement $I(\hat{\rho}_{b:c}:X^{m}) = S(\hat{\rho}_{b:c}) - \sum_{j} p_{j}S(\hat{\rho}_{b:c}^{(j)})$~\cite{Groenewold1971} upper bounded by the Shannon entropy of measurement~\cite{QCQI}. This bound cannot be applied to the process described in the present work for two reasons: we consider the change of internal energy of the battery-charger system as work done on it and the fact that the charger itself functions as a heat bath~\cite{Hovh2020CabT}. 
  Given that the battery-charger initial state is correlated, the thermodynamics of the measurement and feedback process cannot be easily studied.
  
  Nevertheless, it should be expected that the information gain $I(\hat{\rho}_{b:c}:X^{m})$ plays a role in the thermodynamics of measurement and daemonic energy extraction in our model.
  We found it by looking at the dissipated work, which amounts to all the work exerted on a system in a thermodynamic cycle. Given the positive energy $\mathcal{E}_{\rm tot}$ extracted and work $W_{\rm tot}$ done in the thermodynamic cycle, the dissipated work is:

    \begin{eqnarray}\label{WdissGeneral}
        W_{\rm diss} = W_{\rm tot} - \mathcal{E}_{\rm tot} \geq 0,
    \end{eqnarray}
    which must be positive following the second law. We prove this in Appendix \ref{AB}, where we obtain the following lower bound for the dissipated work of a protocol employing a quantum memory:

    \begin{eqnarray}\label{ineqinfo}
        W_{\rm diss} \geq T(H(X^{m})-I(\hat{\rho}_{b:c}:X^{m})),
    \end{eqnarray}
    where we identify the Shannon entropy of measurement $H(X^{m})$, which comes from the reset work of the memory and the information gain $I(\hat{\rho}_{b:c}:X^{m})$. Given that $-I(\hat{\rho}_{b:c}:X^{m})$ is always negative, we can loosely associate it with the daemonic extraction and, thus, to the thermodynamics of the feedback process employed in the protocol.

\begin{figure}
    \includegraphics[scale= 0.53]{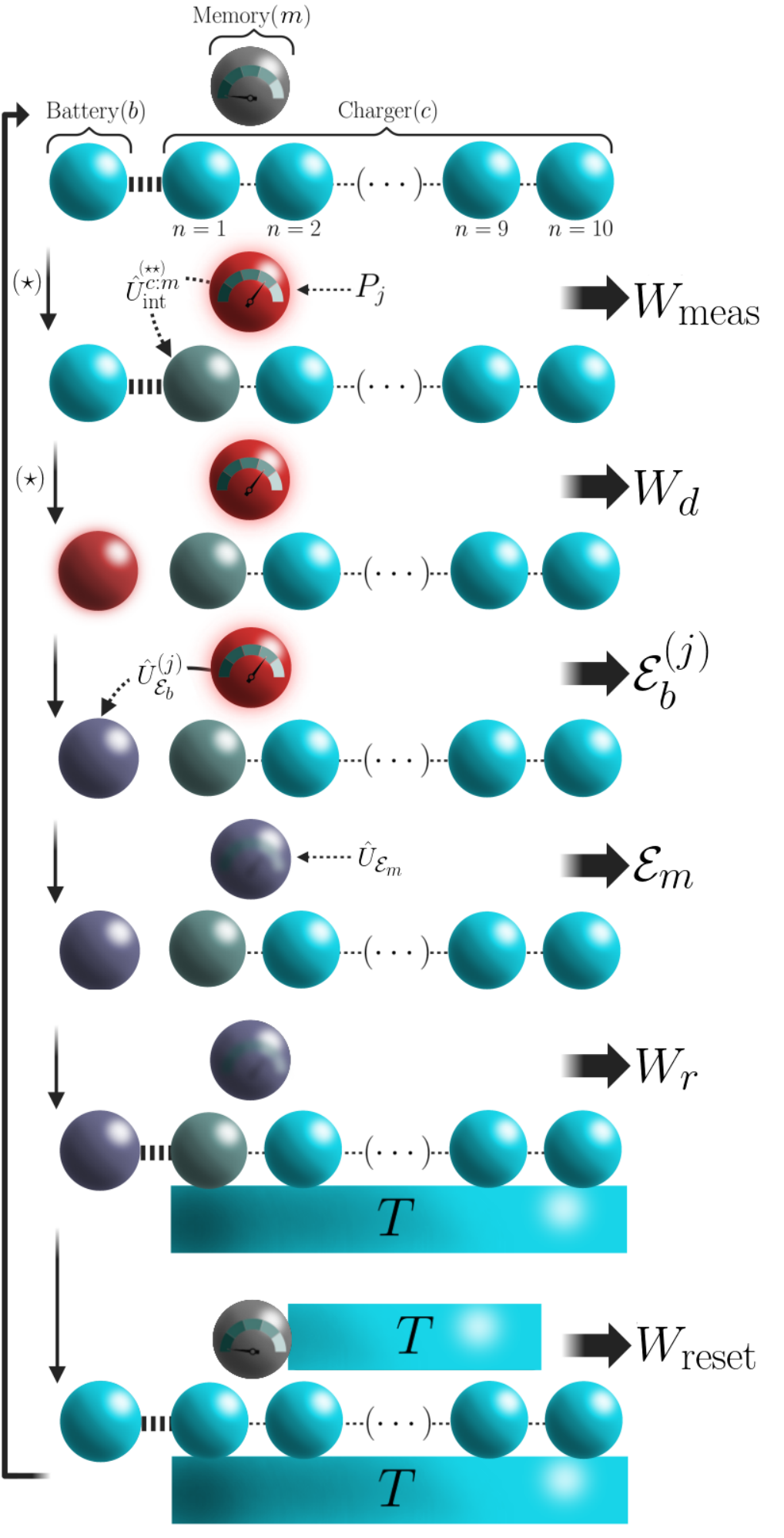}
    \caption{\label{fig:modelo}Scheme of the quantum memory protocol, exemplified by the spin chain model. Each stroke (downward arrow) denotes a step. Steps shown with matching $(\star)$ symbols are interchangeable, though their work costs depend on the ordering. The first stroke executes the measurement on the charger. The second stroke disconnects the battery from the charger. The third stroke daemonically extracts the ergotropy from the battery. The fourth stroke extracts the ergotropy from the memory. The fifth stroke reconnects the battery and charger and thermalizes the joint system. The sixth and final stroke resets the memory to its initial state, and the cycle is ready to start again. The measurement can be done to any $n$ spin of the charger's chain (denoted by the $(\star \star)$ symbol). The plot depicts the case where the measurement is performed on the closest $n=1$ spin to the battery. The numerical results for measuring the closest $n = 1$, and the furthest $n = 10$ spin to the battery are present in the article.}
\end{figure}

    \subsubsection{Comparing two measurements related by a permutation of the measurement operators} \label{Comparison}

The dissipated work defined in Eq. (\ref{WdissGeneral}), can be expressed as
     \begin{eqnarray}
         W_{\rm diss} &=& \text{Tr}[(\hat{H}_{0} + \hat{V}^{b:c}_{\rm int})\sum_{j} p_{j}(\hat{U}_{\mathcal{E}_{b}}^{(j)} \hat{\rho}_{b:c}^{(j)}\hat{U}_{\mathcal{E}_{b}}^{(j)\dag} - \hat{\rho}_{b:c})] \nonumber\\&& + W_{\rm reset}+\Delta E_m,
         \label{Wdiss}
     \end{eqnarray}
where $\Delta E_m$ is the memory's energy in the passive state minus the energy in the standard state.
Appendix \ref{AB} proves this. See Eq. \eqref{ANew}.      
$W_{\rm diss}$ is related to the efficiency $\eta$ by:
    \begin{eqnarray}
        \eta = \frac{\mathcal{E}}{W_{\rm diss} + \mathcal{E}} \leq 1.
    \end{eqnarray} 
    
\begin{figure*}
    \includegraphics[scale= 0.35]{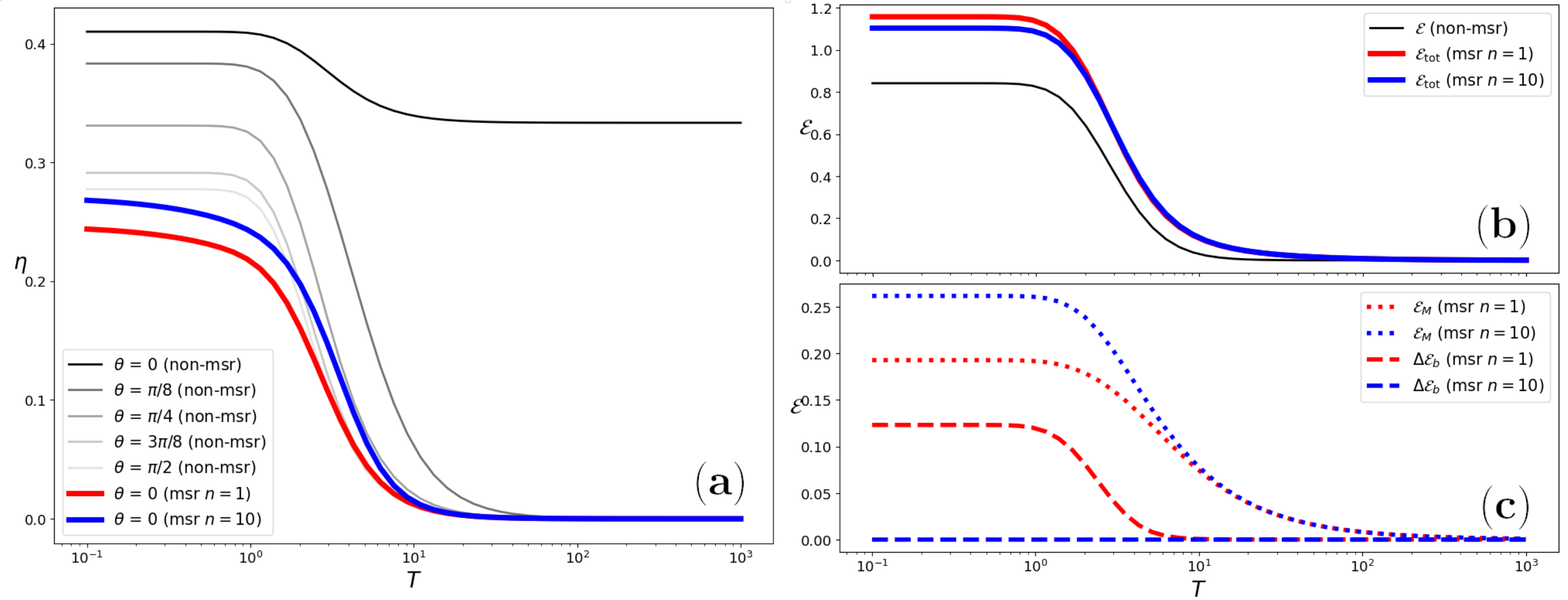}
    \caption{\label{fig:zbasis} $\bf{(a)}$ Efficiency when measuring $n = 1$ (solid red), when measuring $n = 10$ (solid blue) and non-measured cases (grayscale curves). $\bf{(b)}$ Total energy extracted from the protocol when measuring $n = 1$ (solid red), when measuring $n = 10$ (solid blue) and non-measured cases (solid black). $\bf{(c)}$ Memory ergotropy when measuring $n = 1$ (dotted red), when measuring $n = 10$ (dotted blue) and daemonic gain when measuring $n = 1$ (dashed red), when measuring $n = 10$ (dashed blue). $\phi = \pi$ (z-basis of the $n$-th charger's spin), $h_b = 4$, $h_c = 2$, $h_{m} = 0.2$, $\kappa_b = 4$, $\kappa_c = 0.2$.}
\end{figure*}
    
The inequality follows from Eq. (\ref{WdissGeneral}). 
The following observation will be important. Consider a set of measurement operators $\{M_j\}$ associated with the outcomes $j$ and another set of measurement operators $\{M'_j\}$ that is a permutation of the first. Then, measuring with one set or the other will produce the same outcomes but with probabilities and post-measurement states that are a permutation of each other. The quantity $\sum_{j} p_{j}(\hat{U}_{\mathcal{E}_{b}}^{(j)} \hat{\rho}_{b:c}^{(j)}\hat{U}_{\mathcal{E}_{b}}^{(j)\dag} - \hat{\rho}_{b:c})]$ and the passive state of the memory are identical for both measurements and thus, if $W_{\rm reset}$ is the same for measurement $\{M_j\}$ and $\{M'_j\}$, then $W_{\rm diss}$ in Eq.\eqref{Wdiss} is also the same. However, the post measurement state of the battery, which in one case is $\hat{\rho}'_m=\sum_j p_j\ket{m_j}\bra{m_j}$ in the other case will associate to each $\ket{m_j}$ a different probability from the set $\{p_j\}$. This permutation implies that if, after measuring with the first set, the memory is in a passive state, after measuring with the second set, the memory is in an active state from which we can extract its ergotropy $\mathcal{E}_{m}$. Thus, the second protocol is more efficient unless the memory is degenerate, in which case it is always in a passive state.

To summarize, we have $\eta_{+} > \eta_{-} = \eta_{\text{dg}},$
    where $\eta_{(\pm)}$ are the efficiencies of two process with measurement operators $\{M_j\}$ and $\{M_j'\}$ involving a non-degenerate quantum memory with an active (+) or passive (-) post-measurement state. If the memory is degenerate, the efficiency in both cases is $\eta_{\text{dg}}$.

    The next section presents a simple model for implementing a quantum memory with these two sets of measurement operators available.

\section{Transverse Ising spin chain as a daemonic quantum battery model} \label{QMBattery}

We illustrate our results and compare the protocol that uses measurements (section \ref{QMMem}) with the one that does not (section \ref{sec2a}) in the following model. The battery system is a single spin (two-level system) with Hamiltonian $\hat{H}_{b} = -h_b \sigma^{z}_b$, thus the free set $\{ \theta_j \}$ in the ergotropy extracting unitary reduces to a single phase $\theta$. The charger system is an open chain of $N = 10$ spins; the Hamiltonian of the whole chain is given by $\hat{H}_{c} = -(\sum_{n=1} ^{N-1} h_c(\sigma^{z}_{n} + \sigma^{x}_{n}) + \kappa_c \sigma^{x}_{n} \sigma^{x}_{n+1}) - h_c(\sigma^{z}_{N} + \sigma^{x}_{N})$. The interaction between the battery and the charger is given by an xx-spin coupling with strength $\kappa_{b}$ between the battery and the first spin ($n=1$) of the charger, $\hat{V}^{b:c}_{\rm int}= -\kappa_{b} (\sigma^{x}_{b} \sigma^{x}_{1})$. The quantum memory is also a single spin with Hamiltonian $\hat{H}_{m} = -h_{m} \sigma^{z}_{m}$. Figure \ref{fig:modelo} presents a protocol schematic.  

We use the following notation: The eigenstates of the Pauli matrix $-\sigma^{z}_\gamma$ with eigenvalue $ -1$ and $ +1$ are $\ket{0}_\gamma$ and $\ket{1}_\gamma$ respectively. Given this nomenclature, where $\gamma$ stands for $m$, the memory subsystem or $n$, a particular spin of the charger, the initial (and final) state of the memory is $\hat{\rho}_{m} = \ket{0}_m \bra{0}$.

    The memory is coupled to any desired $n$ spin of the charger. Then a projective measurement is done on the memory in the eigenbasis of $\hat{H}_{m}$. This coupling is realized by a (generalized) CNOT gate~\cite{CNOT}, where the control qubit is the charger's $n$-spin to be measured, and the target qubit is the memory. 
    The explicit form of the gate is:

\begin{figure*}
    \includegraphics[scale= 0.35]{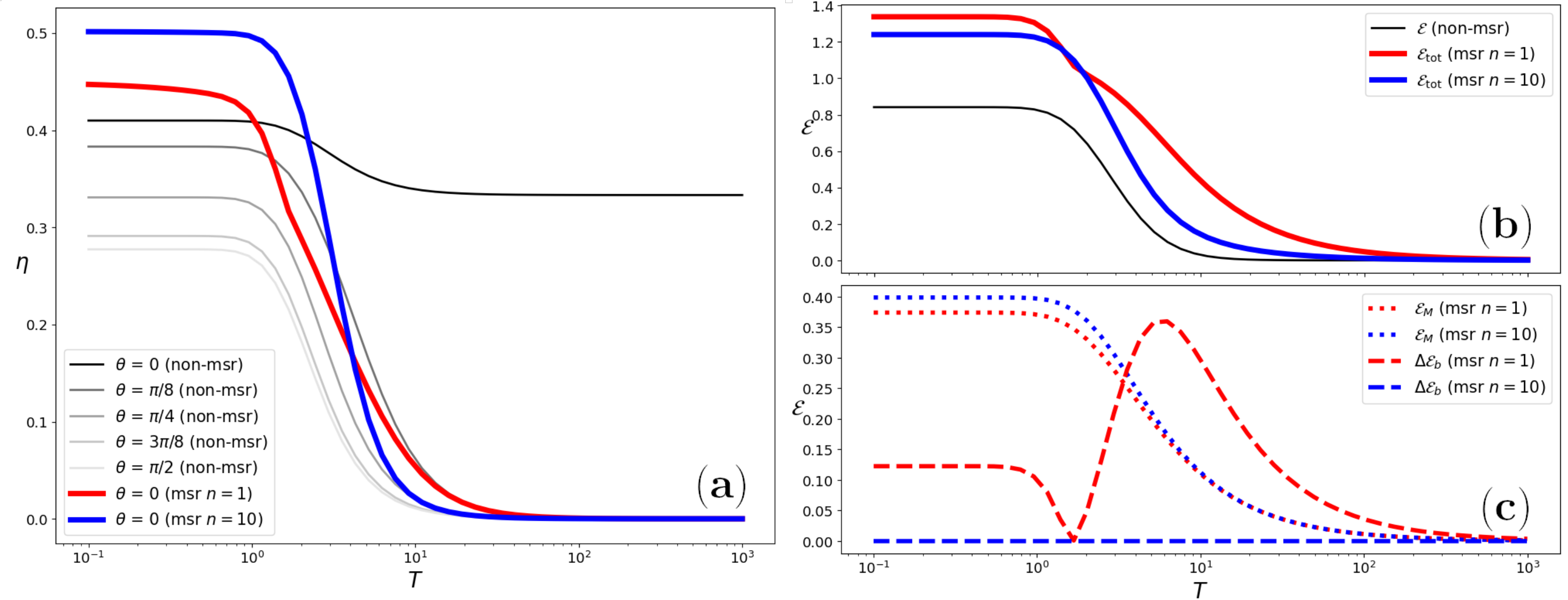}
    \caption{\label{fig:xzbasis4}$\bf{(a)}$ Efficiency when measuring $n = 1$ (solid red), when measuring $n = 10$ (solid blue) and non-measured cases (grayscale curves). $\bf{(b)}$ Total energy extracted from the protocol when measuring $n = 1$ (solid red), when measuring $n = 10$ (solid blue) and non-measured cases (solid black). $\bf{(c)}$ Memory ergotropy when measuring $n = 1$ (dotted red), when measuring $n = 10$ (dotted blue) and daemonic gain when measuring $n = 1$ (dashed red), when measuring $n = 10$ (dashed blue). $\phi = 5\pi/4$ (xz-basis of the $n$-th charger's spin), $h_b = 4$, $h_c = 2$, $h_{m} = 0.2$, $\kappa_b = 4$, $\kappa_c = 0.2$.}
\end{figure*}

    \small
    \begin{eqnarray}
        \text{CNOT}_{n:m}(\phi) &=& e^{-i\sigma^{y}_{n}\phi/2} \ket{0}_{n}\bra{0} e^{i\sigma^{y}_{n}\phi/2} \otimes I_{m} \nonumber\\&& + e^{-i\sigma^{y}_{n}\phi/2}\ket{1}_{n}\bra{1}e^{i\sigma^{y}_{n}\phi/2} \otimes \sigma^{x}_{m}, 
    \end{eqnarray}
    \normalsize
which includes a rotation along the $y$ axis by an angle $\phi$ for the z-basis of the charger's $n$-spin. 
    The outcomes $j\in\{0,1\}$ have probability $p_j(\phi)={\rm Tr}[M_j(\phi)\hat{\rho}_n]$ where $\hat{\rho}_n$ is the reduced state of the $n-$th spin of the charger. The measurement operators are the projectors
    \begin{eqnarray}
        M_j(\phi) = \bra{m_j} \text{CNOT}_{n:m}(\phi)\ket{0}_{m},
    \end{eqnarray}
    and satisfy $M_0(\phi)=M_1(\phi+\pi),M_1(\phi)=M_0(\phi+\pi)$. Thus, we have pairs $\{M_j(\phi),M_j(\phi+\pi)\}$ of measurement operators with the properties discussed in Sec.\ref{Comparison}. The post-measurement states of the memory are $\hat{\rho}_m'(\phi)=p_0(\phi)\ket{0}_m\bra{0}+p_1(\phi)\ket{1}_m\bra{1}$ and $\hat{\rho}_m'(\phi+\pi)=p_1(\phi)\ket{0}_m\bra{0}+p_0(\phi)\ket{1}_m\bra{1}.$ If one is an active state, the other is passive because the populations have been inverted.  Similarly, the reduced state of $n-$th spin after the measurement are $ e^{-i\sigma^{y}_{n}\phi/2}\ket{0}_n \bra{0}e^{i\sigma^{y}_{n}\phi/2}$ for  $M_0(\phi)$ or $M_1(\phi+\pi)$ and the state
    $e^{-i\sigma^{y}_{n}\phi/2}\ket{1}_n \bra{1}e^{i\sigma^{y}_{n}\phi/2}$ for $M_1(\phi)$ or $M_0(\phi+\pi).$   
    Thus, we have a pair of measurement operators from which we extract the same daemonic ergotropy and have the same dissipated work, but one leaves the memory in an active state while the other leaves it in a passive state.  
    
    In the next section, we discuss the energetics of the protocol for different setups. As the reader may have noticed, the protocol we discuss does not fix the memory's resetting process. This process is independent of the post-measurement state of the memory. All our results obtained until now, and in the appendix, are independent of this process. But we have to specify one for the illustration in the next section.  We consider a resetting process that achieves the lower bound for the resetting work, i.e. $W_{\rm reset}=\Delta F_{\rm reset}$, Eq.\eqref{FreeEreset}. For other options, the efficiencies of the measured protocols will be smaller. 

\section{Results} \label{Results}

For the numerical computations we take $h_b = 4$ and $h_{m} = 0.2$. This implies that the maximum ergotropy that the memory can hold is smaller than that of the battery. The charger has $h_c = 2$ and $\kappa_c = 0.2$, so the coupling is small. Considering the nature of the protocol, the coupling strength $\kappa_b$ is considered a control parameter capable of tuning the ergotropy of the battery. All protocols that do not employ measurements were studied for five different phases $\theta = \{ 0, \pi/8, \pi/4, 3\pi/8, \pi/2 \}$. On the other hand, the measured protocols were only studied for the most optimal phase, which was found to be $\theta = 0$ by inspection. We present results for measurements done to the first $n=1$ (closest to the battery) and last $n=10$ (furthest from the battery) spins of the charger's chain. The measurement is executed at the start of the protocol, before the disconnection of the battery from the charger (exactly as shown in Figure \ref{fig:modelo}).

\begin{figure*}
    \includegraphics[scale= 0.35]{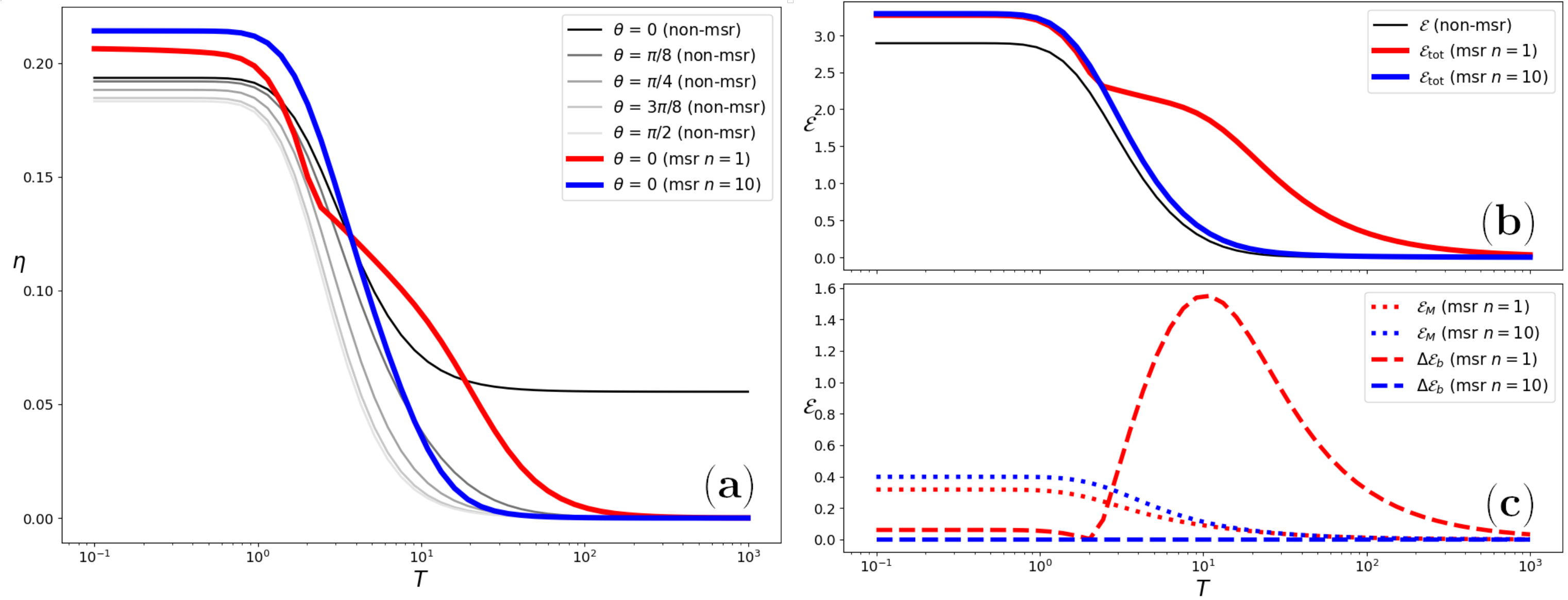}
    \caption{\label{fig:xzbasis16}$\bf{(a)}$ Efficiency when measuring $n = 1$ (solid red), when measuring $n = 10$ (solid blue) and non-measured cases (grayscale curves). $\bf{(b)}$ Total energy extracted from the protocol when measuring $n = 1$ (solid red), when measuring $n = 10$ (solid blue) and non-measured cases (solid black). $\bf{(c)}$ Memory ergotropy when measuring $n = 1$ (dotted red), when measuring $n = 10$ (dotted blue) and daemonic gain when measuring $n = 1$ (dashed red), when measuring $n = 10$ (dashed blue). $\phi = 5\pi/4$ (xz-basis of the $n$-th charger's spin), $h_b = 4$, $h_c = 2$, $h_{m} = 0.2$, $\kappa_b = 16$, $\kappa_c = 0.2$.}
\end{figure*}

The main results are the efficiencies $\eta$ shown in sub-figures $\bf{(a)}$ and the ergotropies and daemonic gains shown in sub-figures $\bf{(b)}$ and $\bf{(c)}$ of Figures \ref{fig:zbasis}, \ref{fig:xzbasis4} and \ref{fig:xzbasis16}. 
Additionally, the work done in measured (only for $n = 1$ cases) and non-measured protocols is presented in Figure \ref{fig:work}.

\subsection{$\phi = \pi$ z-basis measure. Figure \ref{fig:zbasis}} \label{Rz}

We start by analyzing the pair of measurements with $\phi=0$ and $\phi=\pi.$ The post-measurement state of the memory is charged only for $\phi = \pi$; thus, its efficiency is higher. Figure \ref{fig:zbasis}$\bf{a}$ considers $\phi = \pi$.

First, we focus on the $n = 1$ case (red curves). Figure \ref{fig:zbasis}$\bf{c}$ shows the daemonic gain for the battery's ergotropy due to the measurement. It also shows that the memory's ergotropy is non-zero and even greater than the daemonic gain. In combination with the ergotropy extracted from the memory, the total ergotropy (shown in Figure \ref{fig:zbasis}$\bf{b}$) is significantly higher than the unmeasured case. Nevertheless, this gain of energy is heavily counteracted by the increase of work done in the cycle due to the measurement and reset of the memory, this leads to inferior efficiency compared to any unmeasured case. 

The data for measurements of the $n = 10$ site are depicted in blue. The memory's ergotropy (Figure \ref{fig:zbasis}$\bf{c}$) is higher in this than in the $n = 1$ case. At the same time, the daemonic gain is negligible ($\sim 10^{-14}$). The reduction of daemonic ergotropy is expected because the correlation between the battery and the charger's site $n$ decreases as $n$ increases. This leads to a slightly lower total ergotropy than measuring the closest ($n=1$) spin to the battery. However, the efficiency is higher in the $n=10$ than in the $n = 1$ case. When we compare with non-measured cases, we observe that only at temperatures between $1$ and $10$, measuring the spin $n=10$ has higher efficiency than some non-measured setups. 

Figure \ref{fig:work}$\bf{a}$ depicts the works as a function of temperature. The reset work (brown curve) is negative at very low temperatures. So, it diverges negatively in the logarithmic scale. Negative reset work is always counteracted by positive measurement work (orange curve)~\cite{Sagawa2009}.

Next, we explore the energetics of another measurement basis, such that measuring the first spin of the charger is more favorable than measuring the last.

\subsection{$\phi = 5 \pi/4$ xz-basis measure. Figures 3 and 4} \label{Rxz}

 The pair of measurements with $\pi/4$ or $5\pi/4$ corresponds (approximately) to a projection to the eigenbasis of $\hat{H}_{c;n}$. Appendix \ref{rationale} explains why this choice benefits the efficiency. Since the memory's post-measurement state is charged only with $\phi = 5\pi/4$, the efficiency is higher for this value. Figure \ref{fig:xzbasis4} and Figure \ref{fig:xzbasis16} consider $\phi = 5\pi/4$.

We analyze two cases for the battery-charger coupling, $\kappa_{b} = 4$ (Figure \ref{fig:xzbasis4}) and $\kappa_{b} = 16$ (Figure \ref{fig:xzbasis16}).

\begin{figure*}
    \includegraphics[scale= 0.32]{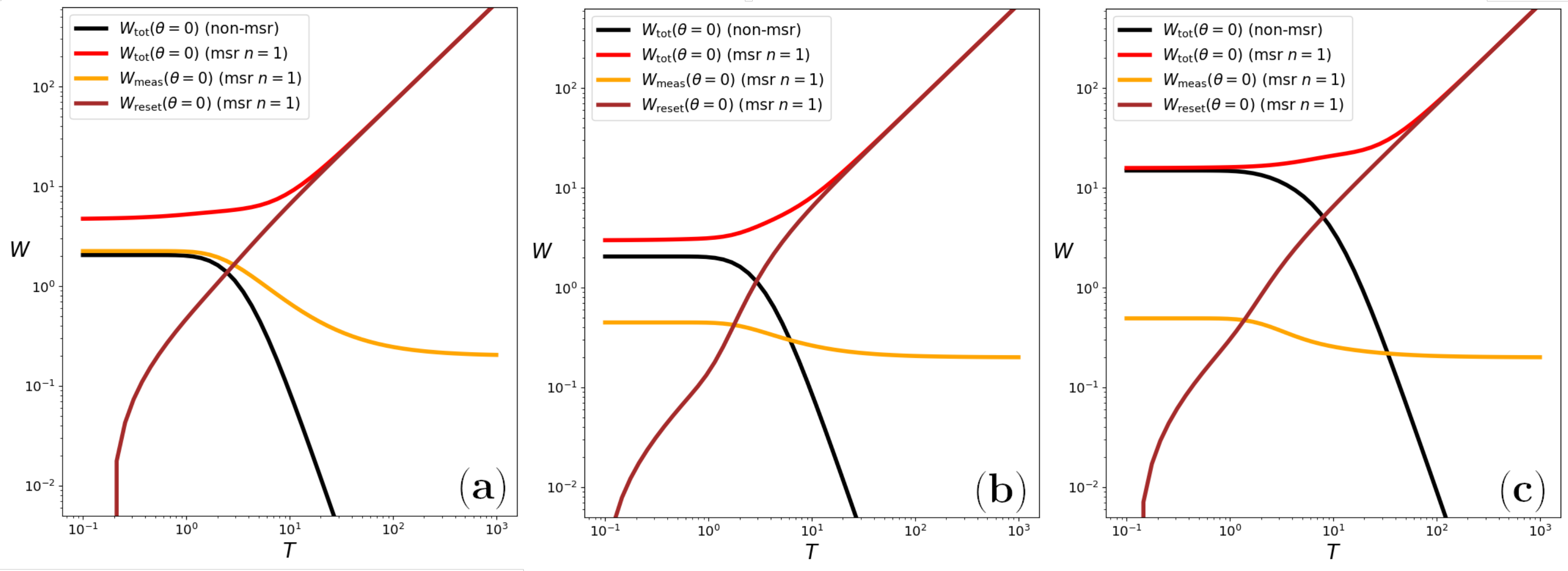}
    \caption{\label{fig:work}Total work done without a measurement (solid black), total work done when measuring $n = 1$ (solid red), measurement work (solid orange) and memory reset work (solid brown) for $h_b = 4$, $h_c = 2$, $h_{m} = 0.2$, $\kappa_c = 0.2$, $\bf{(a)}$ $\phi =\pi$, $\kappa_b = 4$, $\bf{(b)}$ $\phi =5\pi/4$, $\kappa_b = 4$ and $\bf{(c)}$ $\phi =5\pi/4$, $\kappa_b = 16$.}
\end{figure*}

In Figure \ref{fig:xzbasis4}$\bf{b}$ we see that the total ergotropies are slightly higher than the ones for the $\phi = \pi$ case (Figure \ref{fig:zbasis}$\bf{b}$). 
Comparing Figures \ref{fig:work}$\bf{a}$ and \ref{fig:work}$\bf{b}$, we see that the measurement work and the total work when $\phi = 5 \pi /4$ are lower than their counterparts when $\phi = \pi$. So, measuring with $\phi = 5\pi/4$ increases ergotropy and decreases the work cost compared to measuring with $\phi = \pi$, boosting the efficiency enough to surpass all non-measured cases for some temperatures. 

At low temperatures, the efficiency when measuring $n = 10$ is superior to when measuring $n = 1$. For higher temperatures, measuring $n = 1$ gives a slightly superior efficiency than measuring $n = 10$. 
This behavior can be explained by looking at Figures \ref{fig:xzbasis4}$\bf{b}$ and \ref{fig:xzbasis4}$\bf{c}$.

 Figure \ref{fig:xzbasis4}$\bf{b}$ shows that the total ergotropy for $n = 1$ is almost always bigger than for $n = 10$. Furthermore, we notice that the daemonic gain for $n = 1$ (Figure \ref{fig:xzbasis4}$\bf{c}$) has two regimes separated at a characteristic temperature $T_c \sim 2$. 
 
  The memory's ergotropy has no notable behavior other than extracting more energy than the $\phi = \pi$ case. The increase of daemonic gain allows the $n = 1$ case to extract much more energy at higher temperatures than the $n = 10$ case. 

The parameter $\kappa_b$ can be used to tune the ergotropy of the battery. Higher couplings lead to a higher ergotropy (see Figure \ref{fig:xzbasis16}$\bf{b}$), with the trade-off of steep disconnecting/reconnecting work costs as illustrated in Figure \ref{fig:work}$\bf{c}$ (red curve). Figure \ref{fig:xzbasis16}$\bf{a}$ depicts the efficiencies when $\kappa_b = 16$. It behaves similarly to Figure \ref{fig:xzbasis4}$\bf{a}$, except for a noticeable bump in efficiency when measuring $n = 1$. At temperatures around $T = 10$, the efficiency when measuring the $n=1$ charger spin surpasses that of every other protocol. This is due to the particular behavior of the daemonic gain for $n = 1$, shown in Figure \ref{fig:xzbasis16}$\bf{c}$. Though the efficiency around this temperature is only $\eta \sim 0.1$, it extracts considerably more energy than any other protocol at those temperatures. Remarkably, the daemonic gain in Figure \ref{fig:xzbasis16}$\bf{c}$ does not change significantly with increased $\kappa_b$ for temperatures below $T_c$, yet it increases dramatically with $\kappa_b$ above $T_c$. In contrast, the case when $n = 10$ has no notable results, having the same behavior as the previous $n = 10$ cases. 

Finally, it is important to mention that every step in the thermodynamic cycle of the different battery setups has been checked to be consistent with their corresponding bounds presented in this article.

\section{Conclusions} \label{Conclusions}

In this article, we have expanded the scope of the protocol in~\cite{Barra22QBPHT} for quantum batteries charged by thermalization by introducing measurements to the thermodynamic cycle. 

A quantum memory interacts with the charger to execute a quantum measurement. The charger is correlated to the battery, and the measurement information about the battery-charger system is harnessed to increase its energy output. Furthermore, we extract the ergotropy of the memory if it becomes charged by the measurements. The initial state of the battery-charger system is thermal, while the initial state of the memory is pure, called the standard state.

The introduction of memory requires new steps. First, measuring the charger results in an additional work $W_{\rm meas}$ done to the total system. Second, after extracting the memory's ergotropy, it must be reset to its standard state, with an additional work $W_{\rm reset}$. We have shown that although there is a gain in ergotropy, the protocol's overall efficiency improves or worsens compared to the previous one, depending on the measurement and resetting protocol. 

We tested the protocol by considering a battery-charger system consisting of a spin 1/2 chain and a memory consisting of a single spin 1/2 that interacts with a charger's spin. Given the interaction, the measurement results in a projective measurement for the charger's spin. We showed that the measurement basis determines the protocol's efficiency. We found one that, for some model parameters, surpasses the efficiency of the non-measured protocol for a certain range of temperatures.

The protocol in~\cite{Barra22QBPHT} uses the correlations between the battery and charger to decrease the work cost of the cycle. Here, we use them to increase the extracted energy by introducing measurements to the thermodynamic cycle. In both, the protocol's efficiency increases with respect to~\cite{Hovh2020CabT}, but it is found that the one with measurements can surpass the one without.

In summary, our work highlights information as a resource for quantum batteries. It comes with a cost (the measuring and resetting work) but boosts the energy output. The change in efficiency depends on the details of the measurement.

\begin{acknowledgments}
    F. B. Thanks, Fondecyt project 1231210. 
    M. A. S. Acknowledges partial support from Fondecyt project 1231210.
\end{acknowledgments}

\appendix

\section{Proofs of inequalities \eqref{WdissGeneral},\eqref{ineqinfo} and equality \eqref{Wdiss}} \label{AB}

All the names and expressions in this section refer to section \ref{QMMem}.

Recalling the nonequilibrium second law shown in Eq. (\ref{Noneq2law}), a particular result shows that if $\hat{H}^{\prime} = \hat{H}$ and $\hat{\rho}^{\prime}$ is an equilibrium state with respect to $\hat{H}$ at temperature $T$, then $F(\hat{\rho}, \hat{H}) - F(\hat{\rho}^{\prime}, \hat{H}) \geq 0$, which shows that the free energy of a non-equilibrium state is higher than the free energy of its corresponding equilibrium state~\cite{Esposito2011} (the difference reaching zero only when $\hat{\rho} = \hat{\rho}^{\prime}$). Computing the difference of free energies $\Delta F_j$ for our initial equilibrium state $\hat{\rho}_{b:c}$ and final non-equilibrium state $\hat{U}_{{\mathcal E}_b}^{(j)}  \hat{\rho}_{b:c}^{(j)} \hat{U}_{{\mathcal E}_b}^{(j)\dag}$, both with respect to $\hat{H}_{0} +\hat{V}_{\rm int}^{b:c}$, we get:

\begin{widetext}
\begin{eqnarray}
\Delta F_j &=& \text{Tr}[(\hat{H}_{0} + \hat{V}_{\rm int}^{b:c})\hat{U}_{{\mathcal E}_b}^{(j)} \hat{\rho}_{b:c}^{(j)} \hat{U}_{{\mathcal E}_b}^{(j)\dag}] - T S(\hat{U}_{{\mathcal E}_b}^{(j)} \hat{\rho}_{b:c}^{(j)} \hat{U}_{{\mathcal E}_b}^{(j)\dag}) - \text{Tr}[(\hat{H}_{0}+\hat{V}_{\rm int}^{b:c})\hat{\rho}_{b:c}] + T S(\hat{\rho}_{b:c}) \nonumber \\ &=& \text{Tr}[(\hat{H}_{0}+\hat{V}_{\rm int}^{b:c})(\hat{U}_{{\mathcal E}_b}^{(j)} \hat{\rho}_{b:c}^{(j)} \hat{U}_{{\mathcal E}_b}^{(j)\dag} - \hat{\rho}_{b:c})] - T(S(\hat{U}_{{\mathcal E}_b}^{(j)} \hat{\rho}_{b:c}^{(j)} \hat{U}_{{\mathcal E}_b}^{(j)\dag}) - S(\hat{\rho}_{b:c})) \geq 0 \nonumber \\ &\Leftrightarrow& \text{Tr}[(\hat{H}_{0}+\hat{V}_{\rm int}^{b:c})(\hat{U}_{{\mathcal E}_b}^{(j)} \hat{\rho}_{b:c}^{(j)} \hat{U}_{{\mathcal E}_b}^{(j)\dag} - \hat{\rho}_{b:c})] \geq T(S(\hat{U}_{{\mathcal E}_b}^{(j)} \hat{\rho}_{b:c}^{(j)} \hat{U}_{{\mathcal E}_b}^{(j)\dag}) - S(\hat{\rho}_{b:c})) \nonumber \\ &\Leftrightarrow& \sum_{j} p_{j} \text{Tr}[(\hat{H}_{0}+\hat{V}_{\rm int}^{b:c})(\hat{U}_{{\mathcal E}_b}^{(j)} \hat{\rho}_{b:c}^{(j)} \hat{U}_{{\mathcal E}_b}^{(j)\dag} - \hat{\rho}_{b:c})] \geq T(\sum_{j} p_{j} S(\hat{\rho}_{b:c}^{(j)}) - S(\hat{\rho}_{b:c})) ,
 \label{eq:wideeq}
\end{eqnarray}
\end{widetext}
\noindent where in the last inequality we have taken the mean of all possible outcomes and used the fact that the operators that extract the ergotropy leave the entropy of the state intact  $S(\hat{U}_{{\mathcal E}_b}^{(j)} \hat{\rho}_{b:c}^{(j)} \hat{U}_{{\mathcal E}_b}^{(j)\dag}) = S(\hat{\rho}_{b:c}^{(j)})$.

We consider the total dissipated work 
\begin{equation}
    W_{\rm diss} = W_{\rm tot} - \mathcal{E}_{\rm tot}.
\end{equation}

The sum of the disconnecting work and measurement work is $W_{d} + W_{\text{meas}} = \sum_{j} p_{j} (\text{Tr}[(\hat{H}_{0} + \hat{H}_{m})\hat{\rho}_{\rm tot}^{(j)}] - \text{Tr}[\hat{H}_{\rm tot} \hat{\rho}_{\rm tot}])$ regardless of the order of events. Adding the resetting $W_{\rm reset}$ and reconnecting $W_r$ [Eq.\eqref{wreco}] works, and substracting the total ergotropy given in Eq.\eqref{totalergo}, we write the dissipated work explicitly:

\begin{widetext}
\begin{eqnarray}
    W_{\rm diss}  &=& \cancel{\sum_{j} p_{j} \text{Tr}[(\hat{H}_{0} + \hat{H}_{m})\hat{\rho}_{\rm tot}^{(j)}]} - \text{Tr}[\hat{H}_{\rm tot} \hat{\rho}_{\rm tot}] + \sum_{j} p_{j} \text{Tr}[(\hat{H}_{0} + \hat{H}_{m})\hat{U}_{{\mathcal E}_b}^{(j)} \hat{U}_{{\mathcal E}_m}\hat{\rho}_{\rm tot}^{(j)}\hat{U}_{{\mathcal E}_b}^{(j)\dag} \hat{U}_{{\mathcal E}_m}^{\dag}] \nonumber \\ &&  - \cancel{\sum_{j} p_{j} \text{Tr}[(\hat{H}_{0} + \hat{H}_{m}) \hat{\rho}_{\rm tot}^{(j)}]} + \sum_{j} p_{j}\text{Tr}[\hat{V}_{\rm int}^{b:c}\hat{U}_{{\mathcal E}_b}^{(j)} \hat{U}_{{\mathcal E}_m}\hat{\rho}_{\rm tot}^{(j)}\hat{U}_{{\mathcal E}_b}^{(j)\dag} \hat{U}_{{\mathcal E}_m}^{\dag}] + W_{\rm reset} \nonumber 
    \end{eqnarray}

     \begin{eqnarray}\label{ANew}
    W_{\rm diss}&=& \sum_{j} p_{j}\text{Tr}[(\hat{H}_{0} + \hat{V}^{b:c}_{\rm int})(\hat{U}_{{\mathcal E}_b}^{(j)} \hat{\rho}_{b:c}^{(j)}\hat{U}_{{\mathcal E}_b}^{(j)\dag} - \hat{\rho}_{b:c})] +W_{\rm reset}- \text{Tr}[\hat{H}_{m} \hat{\rho}_{m}^{0}]+ \text{Tr}[\hat{H}_{m} \hat{U}_{{\mathcal E}_m} \hat{\rho}_{m}^{\prime}\hat{U}_{{\mathcal E}_m}^{\dag}].
    \end{eqnarray}
    For a degenerate memory, the last two terms cancel. Eq.\eqref{ANew} is Eq.\eqref{Wdiss} in the main text. From this, we will deduce the inequalities. 
    In general, from Eq.\eqref{FreeEreset} 
    \begin{eqnarray}
    W_{\rm reset}&\geq&\text{Tr}[\hat{H}_{m} \hat{\rho}_{m}^{0}] - TS(\hat{\rho}_{m}^{0}) - \text{Tr}[\hat{H}_{m} \hat{U}_{{\mathcal E}_m} \hat{\rho}_{m}^{\prime}\hat{U}_{{\mathcal E}_m}^{\dag}] + TS(\hat{U}_{{\mathcal E}_m} \hat{\rho}_{m}^{\prime}\hat{U}_{{\mathcal E}_m}^{\dag})\nonumber\\
     W_{\rm reset}&-& \text{Tr}[\hat{H}_{m} \hat{\rho}_{m}^{0}]+ \text{Tr}[\hat{H}_{m} \hat{U}_{{\mathcal E}_m} \hat{\rho}_{m}^{\prime}\hat{U}_{{\mathcal E}_m}^{\dag}]\geq TS(\hat{U}_{{\mathcal E}_m} \hat{\rho}_{m}^{\prime}\hat{U}_{{\mathcal E}_m}^{\dag})- TS(\hat{\rho}_{m}^{0})= TH(X^{m})\label{ineq2}
    \end{eqnarray}
    where we used that $S(\hat{\rho}_{m}^{0}) = 0$  because $\hat{\rho}_{m}^{0}$ is a pure state and that $S(\hat{U}_{{\mathcal E}_m} \hat{\rho}_{m}^{\prime}\hat{U}_{{\mathcal E}_m}^{\dag}) = S(\hat{\rho}_{m}^{\prime}) = H(X^{m})$.
    
    Putting together inequalities \eqref{eq:wideeq} and \eqref{ineq2} in Eq.\eqref{ANew}, we obtain
 \begin{eqnarray}
    W_{\rm diss}  \geq  T(-I(\hat{\rho}_{b:c}:X^{m}) + H(X^m)) \nonumber,
\end{eqnarray}
which is inequality \eqref{ineqinfo}.
Remembering that $H(X^{m}) \geq I(\hat{\rho}_{b:c}:X^{m})$ ~\cite{QCQI} we obtain our final result:

\begin{equation}
   W_{\rm diss} \geq 0 \ _\square.
\end{equation}

This proves \eqref{WdissGeneral}, i.e., the positivity of the dissipated work.   
    
\end{widetext}

Concerning the pair of measurements $M_j$ and $M_j'$, which are permutations of each other. In the last term of Eq.\eqref{ANew}, the state $\hat{\rho}_{m}^{\prime} = \sum_{j} p_{j} \ket{m_j}_{m}\bra{m_j}$ is the post-measurement state of the memory. If it is passive, then $p_1\geq p_2\geq\ldots$
and $\hat{U}_{{\mathcal E}_m}$ is the identity. If $\hat{\rho}_{m}^{\prime}$ is active, then $\hat{U}_{{\mathcal E}_m}$ bring it to the passive state. Therefore, if $W_{\rm reset}$ is the same for an active or passive memory, the $W_{\rm diss}$ is also the same. For example, if $W_{\rm reset}$ is give by the lower bound in Eq.\eqref{FreeEreset}, then

\begin{eqnarray}\label{A6}
    W_{\rm diss} &=& \sum_{j} p_{j}\text{Tr}[(\hat{H}_{0} + \hat{V}^{b:c}_{\rm int})(\hat{U}_{\mathcal{E}_{b}}^{(j)} \hat{\rho}_{b:c}^{(j)}\hat{U}_{\mathcal{E}_{b}}^{(j)\dag} - \hat{\rho}_{b:c})] \nonumber\\&& + TH(X^{m}),
\end{eqnarray}
for both passive and active memories.

\section{Rationale for measuring the CNOT with $\phi=5\pi/4$}\label{rationale}
We note that
the work cost needed to measure and disconnect the battery can be reduced to $W_{\rm meas} + W_d = -\text{Tr}[\hat{V}_{\rm int}^{b:c}\rho_{b:c}] + \sum_j p_j \text{Tr}[\hat{H}_c( \hat{\rho}_c^{(j)} - \hat{\rho}_{c})] + \text{Tr}[\hat{H}_{m}(\hat{\rho}_{m}^{\prime} - \hat{\rho}_{m})]$ (see Appendix \ref{AB}). First, from this result, we can see that $-\text{Tr}[\hat{V}_{\rm int}^{b:c}\rho_{b:c}]$ is independent of the measurement protocol and that $\text{Tr}[\hat{H}_{m}(\hat{\rho}_{m}^{\prime} - \hat{\rho}_{m})]$ is always positive, as it reflects the charging of the memory, thus we are left with the measurement-dependent term $\Delta E_c = \sum_j p_j \text{Tr}[\hat{H}_c(\hat{\rho}_c^{(j)} - \hat{\rho}_{c})]$, which is not necessarily positive. Inspired by this result, we now explore the case where $\Delta E_c$ approaches 0, such that the measurement has a minimal effect (on average) on the internal energy of the charger. Such change of internal energy of the charger reduces to 0 when the measurement is a projection in the eigenbasis of $\hat{H}_c$. Given that our measurement protocol only measures a single spin of the charger, we present the following setup that reduces the change of internal energy of the charger to a small value. First, we take the internal coupling of the charger to be small (in the examples shown in this work, $\kappa_c = 0.1 h_c$), we then measure the eigenbasis of the internal Hamiltonian of the $n$-spin of the charger $\hat{H}_{c; n} = -h_c (\sigma_n^{z} + \sigma_n^{x})$ of interest. This basis is a rotation in $\phi=\pi/4$ or $\phi=5\pi/4$ of the eigenbasis of $\sigma_n^z.$

\bibliography{apssamp}

\providecommand{\noopsort}[1]{}\providecommand{\singleletter}[1]{#1}%
\begin{thebibliography}{50}%
\makeatletter
\providecommand \@ifxundefined [1]{%
 \@ifx{#1\undefined}
}%
\providecommand \@ifnum [1]{%
 \ifnum #1\expandafter \@firstoftwo
 \else \expandafter \@secondoftwo
 \fi
}%
\providecommand \@ifx [1]{%
 \ifx #1\expandafter \@firstoftwo
 \else \expandafter \@secondoftwo
 \fi
}%
\providecommand \natexlab [1]{#1}%
\providecommand \enquote  [1]{``#1''}%
\providecommand \bibnamefont  [1]{#1}%
\providecommand \bibfnamefont [1]{#1}%
\providecommand \citenamefont [1]{#1}%
\providecommand \href@noop [0]{\@secondoftwo}%
\providecommand \href [0]{\begingroup \@sanitize@url \@href}%
\providecommand \@href[1]{\@@startlink{#1}\@@href}%
\providecommand \@@href[1]{\endgroup#1\@@endlink}%
\providecommand \@sanitize@url [0]{\catcode `\\12\catcode `\$12\catcode
  `\&12\catcode `\#12\catcode `\^12\catcode `\_12\catcode `\%12\relax}%
\providecommand \@@startlink[1]{}%
\providecommand \@@endlink[0]{}%
\providecommand \url  [0]{\begingroup\@sanitize@url \@url }%
\providecommand \@url [1]{\endgroup\@href {#1}{\urlprefix }}%
\providecommand \urlprefix  [0]{URL }%
\providecommand \Eprint [0]{\href }%
\providecommand \doibase [0]{https://doi.org/}%
\providecommand \selectlanguage [0]{\@gobble}%
\providecommand \bibinfo  [0]{\@secondoftwo}%
\providecommand \bibfield  [0]{\@secondoftwo}%
\providecommand \translation [1]{[#1]}%
\providecommand \BibitemOpen [0]{}%
\providecommand \bibitemStop [0]{}%
\providecommand \bibitemNoStop [0]{.\EOS\space}%
\providecommand \EOS [0]{\spacefactor3000\relax}%
\providecommand \BibitemShut  [1]{\csname bibitem#1\endcsname}%
\let\auto@bib@innerbib\@empty
\bibitem [{\citenamefont {Alicki}\ and\ \citenamefont
  {Fannes}(2013)}]{Alicki2013}%
  \BibitemOpen
  \bibfield  {author} {\bibinfo {author} {\bibfnamefont {R.}~\bibnamefont
  {Alicki}}\ and\ \bibinfo {author} {\bibfnamefont {M.}~\bibnamefont
  {Fannes}},\ }\bibfield  {title} {\bibinfo {title} {Entanglement boost for
  extractable work from ensembles of quantum batteries},\ }\href
  {https://doi.org/10.1103/PhysRevE.87.042123} {\bibfield  {journal} {\bibinfo
  {journal} {Phys. Rev. E}\ }\textbf {\bibinfo {volume} {87}},\ \bibinfo
  {pages} {042123} (\bibinfo {year} {2013})}\BibitemShut {NoStop}%
\bibitem [{\citenamefont {Binder}\ \emph {et~al.}(2015)\citenamefont {Binder},
  \citenamefont {Vinjanampathy}, \citenamefont {Modi},\ and\ \citenamefont
  {Goold}}]{Binder2015}%
  \BibitemOpen
  \bibfield  {author} {\bibinfo {author} {\bibfnamefont {F.~C.}\ \bibnamefont
  {Binder}}, \bibinfo {author} {\bibfnamefont {S.}~\bibnamefont
  {Vinjanampathy}}, \bibinfo {author} {\bibfnamefont {K.}~\bibnamefont
  {Modi}},\ and\ \bibinfo {author} {\bibfnamefont {J.}~\bibnamefont {Goold}},\
  }\bibfield  {title} {\bibinfo {title} {Quantacell: powerful charging of
  quantum batteries},\ }\href {https://doi.org/10.1088/1367-2630/17/7/075015}
  {\bibfield  {journal} {\bibinfo  {journal} {New Journal of Physics}\ }\textbf
  {\bibinfo {volume} {17}},\ \bibinfo {pages} {075015} (\bibinfo {year}
  {2015})}\BibitemShut {NoStop}%
\bibitem [{\citenamefont {Barra}(2019)}]{Barra2019}%
  \BibitemOpen
  \bibfield  {author} {\bibinfo {author} {\bibfnamefont {F.}~\bibnamefont
  {Barra}},\ }\bibfield  {title} {\bibinfo {title} {Dissipative charging of a
  quantum battery},\ }\href {https://doi.org/10.1103/PhysRevLett.122.210601}
  {\bibfield  {journal} {\bibinfo  {journal} {Phys. Rev. Lett.}\ }\textbf
  {\bibinfo {volume} {122}},\ \bibinfo {pages} {210601} (\bibinfo {year}
  {2019})}\BibitemShut {NoStop}%
\bibitem [{\citenamefont {Farina}\ \emph {et~al.}(2019)\citenamefont {Farina},
  \citenamefont {Andolina}, \citenamefont {Mari}, \citenamefont {Polini},\ and\
  \citenamefont {Giovannetti}}]{Farina2019}%
  \BibitemOpen
  \bibfield  {author} {\bibinfo {author} {\bibfnamefont {D.}~\bibnamefont
  {Farina}}, \bibinfo {author} {\bibfnamefont {G.~M.}\ \bibnamefont
  {Andolina}}, \bibinfo {author} {\bibfnamefont {A.}~\bibnamefont {Mari}},
  \bibinfo {author} {\bibfnamefont {M.}~\bibnamefont {Polini}},\ and\ \bibinfo
  {author} {\bibfnamefont {V.}~\bibnamefont {Giovannetti}},\ }\bibfield
  {title} {\bibinfo {title} {Charger-mediated energy transfer for quantum
  batteries: An open-system approach},\ }\href
  {https://doi.org/10.1103/PhysRevB.99.035421} {\bibfield  {journal} {\bibinfo
  {journal} {Phys. Rev. B}\ }\textbf {\bibinfo {volume} {99}},\ \bibinfo
  {pages} {035421} (\bibinfo {year} {2019})}\BibitemShut {NoStop}%
\bibitem [{\citenamefont {Pirmoradian}\ and\ \citenamefont
  {M\o{}lmer}(2019)}]{Pirmoradian2019}%
  \BibitemOpen
  \bibfield  {author} {\bibinfo {author} {\bibfnamefont {F.}~\bibnamefont
  {Pirmoradian}}\ and\ \bibinfo {author} {\bibfnamefont {K.}~\bibnamefont
  {M\o{}lmer}},\ }\bibfield  {title} {\bibinfo {title} {Aging of a quantum
  battery},\ }\href {https://doi.org/10.1103/PhysRevA.100.043833} {\bibfield
  {journal} {\bibinfo  {journal} {Phys. Rev. A}\ }\textbf {\bibinfo {volume}
  {100}},\ \bibinfo {pages} {043833} (\bibinfo {year} {2019})}\BibitemShut
  {NoStop}%
\bibitem [{\citenamefont {Santos}\ \emph {et~al.}(2019)\citenamefont {Santos},
  \citenamefont {\ifmmode~\mbox{\c{C}}\else \c{C}\fi{}akmak}, \citenamefont
  {Campbell},\ and\ \citenamefont {Zinner}}]{stableadiabatic}%
  \BibitemOpen
  \bibfield  {author} {\bibinfo {author} {\bibfnamefont {A.~C.}\ \bibnamefont
  {Santos}}, \bibinfo {author} {\bibfnamefont {B.~i. e. i. f. m.~c.}\
  \bibnamefont {\ifmmode~\mbox{\c{C}}\else \c{C}\fi{}akmak}}, \bibinfo {author}
  {\bibfnamefont {S.}~\bibnamefont {Campbell}},\ and\ \bibinfo {author}
  {\bibfnamefont {N.~T.}\ \bibnamefont {Zinner}},\ }\bibfield  {title}
  {\bibinfo {title} {Stable adiabatic quantum batteries},\ }\href
  {https://doi.org/10.1103/PhysRevE.100.032107} {\bibfield  {journal} {\bibinfo
   {journal} {Phys. Rev. E}\ }\textbf {\bibinfo {volume} {100}},\ \bibinfo
  {pages} {032107} (\bibinfo {year} {2019})}\BibitemShut {NoStop}%
\bibitem [{\citenamefont {Hovhannisyan}\ \emph {et~al.}(2020)\citenamefont
  {Hovhannisyan}, \citenamefont {Barra},\ and\ \citenamefont
  {Imparato}}]{Hovh2020CabT}%
  \BibitemOpen
  \bibfield  {author} {\bibinfo {author} {\bibfnamefont {K.~V.}\ \bibnamefont
  {Hovhannisyan}}, \bibinfo {author} {\bibfnamefont {F.}~\bibnamefont
  {Barra}},\ and\ \bibinfo {author} {\bibfnamefont {A.}~\bibnamefont
  {Imparato}},\ }\bibfield  {title} {\bibinfo {title} {Charging assisted by
  thermalization},\ }\href {https://doi.org/10.1103/PhysRevResearch.2.033413}
  {\bibfield  {journal} {\bibinfo  {journal} {Phys. Rev. Res.}\ }\textbf
  {\bibinfo {volume} {2}},\ \bibinfo {pages} {033413} (\bibinfo {year}
  {2020})}\BibitemShut {NoStop}%
\bibitem [{\citenamefont {Zhao}\ \emph {et~al.}(2021)\citenamefont {Zhao},
  \citenamefont {Dou},\ and\ \citenamefont {Zhao}}]{Zhao2021}%
  \BibitemOpen
  \bibfield  {author} {\bibinfo {author} {\bibfnamefont {F.}~\bibnamefont
  {Zhao}}, \bibinfo {author} {\bibfnamefont {F.-Q.}\ \bibnamefont {Dou}},\ and\
  \bibinfo {author} {\bibfnamefont {Q.}~\bibnamefont {Zhao}},\ }\bibfield
  {title} {\bibinfo {title} {Quantum battery of interacting spins with
  environmental noise},\ }\href {https://doi.org/10.1103/PhysRevA.103.033715}
  {\bibfield  {journal} {\bibinfo  {journal} {Phys. Rev. A}\ }\textbf {\bibinfo
  {volume} {103}},\ \bibinfo {pages} {033715} (\bibinfo {year}
  {2021})}\BibitemShut {NoStop}%
\bibitem [{\citenamefont {Qi}\ and\ \citenamefont {Jing}(2021)}]{Jing2021}%
  \BibitemOpen
  \bibfield  {author} {\bibinfo {author} {\bibfnamefont {S.-f.}\ \bibnamefont
  {Qi}}\ and\ \bibinfo {author} {\bibfnamefont {J.}~\bibnamefont {Jing}},\
  }\bibfield  {title} {\bibinfo {title} {Magnon-mediated quantum battery under
  systematic errors},\ }\href {https://doi.org/10.1103/PhysRevA.104.032606}
  {\bibfield  {journal} {\bibinfo  {journal} {Phys. Rev. A}\ }\textbf {\bibinfo
  {volume} {104}},\ \bibinfo {pages} {032606} (\bibinfo {year}
  {2021})}\BibitemShut {NoStop}%
\bibitem [{\citenamefont {Carrasco}\ \emph {et~al.}(2022)\citenamefont
  {Carrasco}, \citenamefont {Maze}, \citenamefont {Hermann-Avigliano},\ and\
  \citenamefont {Barra}}]{Carrasco2022}%
  \BibitemOpen
  \bibfield  {author} {\bibinfo {author} {\bibfnamefont {J.}~\bibnamefont
  {Carrasco}}, \bibinfo {author} {\bibfnamefont {J.~R.}\ \bibnamefont {Maze}},
  \bibinfo {author} {\bibfnamefont {C.}~\bibnamefont {Hermann-Avigliano}},\
  and\ \bibinfo {author} {\bibfnamefont {F.}~\bibnamefont {Barra}},\ }\bibfield
   {title} {\bibinfo {title} {Collective enhancement in dissipative quantum
  batteries},\ }\href {https://doi.org/10.1103/PhysRevE.105.064119} {\bibfield
  {journal} {\bibinfo  {journal} {Phys. Rev. E}\ }\textbf {\bibinfo {volume}
  {105}},\ \bibinfo {pages} {064119} (\bibinfo {year} {2022})}\BibitemShut
  {NoStop}%
\bibitem [{\citenamefont {Yao}\ and\ \citenamefont {Shao}(2022)}]{Yao2022}%
  \BibitemOpen
  \bibfield  {author} {\bibinfo {author} {\bibfnamefont {Y.}~\bibnamefont
  {Yao}}\ and\ \bibinfo {author} {\bibfnamefont {X.~Q.}\ \bibnamefont {Shao}},\
  }\bibfield  {title} {\bibinfo {title} {Optimal charging of open spin-chain
  quantum batteries via homodyne-based feedback control},\ }\href
  {https://doi.org/10.1103/PhysRevE.106.014138} {\bibfield  {journal} {\bibinfo
   {journal} {Phys. Rev. E}\ }\textbf {\bibinfo {volume} {106}},\ \bibinfo
  {pages} {014138} (\bibinfo {year} {2022})}\BibitemShut {NoStop}%
\bibitem [{\citenamefont {Hu}\ \emph {et~al.}(2022)\citenamefont {Hu},
  \citenamefont {Qiu}, \citenamefont {Souza}, \citenamefont {Yuan},
  \citenamefont {Zhou}, \citenamefont {Zhang}, \citenamefont {Chu},
  \citenamefont {Pan}, \citenamefont {Hu}, \citenamefont {Li}, \citenamefont
  {Xu}, \citenamefont {Zhong}, \citenamefont {Liu}, \citenamefont {Yan},
  \citenamefont {Tan}, \citenamefont {Bachelard}, \citenamefont {Villas-Boas},
  \citenamefont {Santos},\ and\ \citenamefont {Yu}}]{Hu_2022}%
  \BibitemOpen
  \bibfield  {author} {\bibinfo {author} {\bibfnamefont {C.-K.}\ \bibnamefont
  {Hu}}, \bibinfo {author} {\bibfnamefont {J.}~\bibnamefont {Qiu}}, \bibinfo
  {author} {\bibfnamefont {P.~J.~P.}\ \bibnamefont {Souza}}, \bibinfo {author}
  {\bibfnamefont {J.}~\bibnamefont {Yuan}}, \bibinfo {author} {\bibfnamefont
  {Y.}~\bibnamefont {Zhou}}, \bibinfo {author} {\bibfnamefont {L.}~\bibnamefont
  {Zhang}}, \bibinfo {author} {\bibfnamefont {J.}~\bibnamefont {Chu}}, \bibinfo
  {author} {\bibfnamefont {X.}~\bibnamefont {Pan}}, \bibinfo {author}
  {\bibfnamefont {L.}~\bibnamefont {Hu}}, \bibinfo {author} {\bibfnamefont
  {J.}~\bibnamefont {Li}}, \bibinfo {author} {\bibfnamefont {Y.}~\bibnamefont
  {Xu}}, \bibinfo {author} {\bibfnamefont {Y.}~\bibnamefont {Zhong}}, \bibinfo
  {author} {\bibfnamefont {S.}~\bibnamefont {Liu}}, \bibinfo {author}
  {\bibfnamefont {F.}~\bibnamefont {Yan}}, \bibinfo {author} {\bibfnamefont
  {D.}~\bibnamefont {Tan}}, \bibinfo {author} {\bibfnamefont {R.}~\bibnamefont
  {Bachelard}}, \bibinfo {author} {\bibfnamefont {C.~J.}\ \bibnamefont
  {Villas-Boas}}, \bibinfo {author} {\bibfnamefont {A.~C.}\ \bibnamefont
  {Santos}},\ and\ \bibinfo {author} {\bibfnamefont {D.}~\bibnamefont {Yu}},\
  }\bibfield  {title} {\bibinfo {title} {Optimal charging of a superconducting
  quantum battery},\ }\href {https://doi.org/10.1088/2058-9565/ac8444}
  {\bibfield  {journal} {\bibinfo  {journal} {Quantum Science and Technology}\
  }\textbf {\bibinfo {volume} {7}},\ \bibinfo {pages} {045018} (\bibinfo {year}
  {2022})}\BibitemShut {NoStop}%
\bibitem [{\citenamefont {Barra}\ \emph {et~al.}(2022)\citenamefont {Barra},
  \citenamefont {Hovhannisyan},\ and\ \citenamefont {Imparato}}]{Barra22QBPHT}%
  \BibitemOpen
  \bibfield  {author} {\bibinfo {author} {\bibfnamefont {F.}~\bibnamefont
  {Barra}}, \bibinfo {author} {\bibfnamefont {K.~V.}\ \bibnamefont
  {Hovhannisyan}},\ and\ \bibinfo {author} {\bibfnamefont {A.}~\bibnamefont
  {Imparato}},\ }\bibfield  {title} {\bibinfo {title} {Quantum batteries at the
  verge of a phase transition},\ }\href
  {https://doi.org/10.1088/1367-2630/ac43ed} {\bibfield  {journal} {\bibinfo
  {journal} {New J. Phys.}\ }\textbf {\bibinfo {volume} {24}},\ \bibinfo
  {pages} {015003} (\bibinfo {year} {2022})}\BibitemShut {NoStop}%
\bibitem [{\citenamefont {Shaghaghi}\ \emph {et~al.}(2022)\citenamefont
  {Shaghaghi}, \citenamefont {Singh}, \citenamefont {Benenti},\ and\
  \citenamefont {Rosa}}]{micromaser1}%
  \BibitemOpen
  \bibfield  {author} {\bibinfo {author} {\bibfnamefont {V.}~\bibnamefont
  {Shaghaghi}}, \bibinfo {author} {\bibfnamefont {V.}~\bibnamefont {Singh}},
  \bibinfo {author} {\bibfnamefont {G.}~\bibnamefont {Benenti}},\ and\ \bibinfo
  {author} {\bibfnamefont {D.}~\bibnamefont {Rosa}},\ }\bibfield  {title}
  {\bibinfo {title} {Micromasers as quantum batteries},\ }\href
  {https://doi.org/10.1088/2058-9565/ac8829} {\bibfield  {journal} {\bibinfo
  {journal} {Quantum Science and Technology}\ }\textbf {\bibinfo {volume}
  {7}},\ \bibinfo {pages} {04LT01} (\bibinfo {year} {2022})}\BibitemShut
  {NoStop}%
\bibitem [{\citenamefont {Morrone}\ \emph
  {et~al.}(2023{\natexlab{a}})\citenamefont {Morrone}, \citenamefont {Rossi},
  \citenamefont {Smirne},\ and\ \citenamefont {Genoni}}]{Morrone_2023}%
  \BibitemOpen
  \bibfield  {author} {\bibinfo {author} {\bibfnamefont {D.}~\bibnamefont
  {Morrone}}, \bibinfo {author} {\bibfnamefont {M.~A.~C.}\ \bibnamefont
  {Rossi}}, \bibinfo {author} {\bibfnamefont {A.}~\bibnamefont {Smirne}},\ and\
  \bibinfo {author} {\bibfnamefont {M.~G.}\ \bibnamefont {Genoni}},\ }\bibfield
   {title} {\bibinfo {title} {Charging a quantum battery in a non-markovian
  environment: a collisional model approach},\ }\href
  {https://doi.org/10.1088/2058-9565/accca4} {\bibfield  {journal} {\bibinfo
  {journal} {Quantum Science and Technology}\ }\textbf {\bibinfo {volume}
  {8}},\ \bibinfo {pages} {035007} (\bibinfo {year}
  {2023}{\natexlab{a}})}\BibitemShut {NoStop}%
\bibitem [{\citenamefont {Ge}\ \emph {et~al.}(2023)\citenamefont {Ge},
  \citenamefont {Yu}, \citenamefont {Xin}, \citenamefont {Wang}, \citenamefont
  {Zhang}, \citenamefont {Zheng}, \citenamefont {Li}, \citenamefont {Lan},\
  and\ \citenamefont {Yu}}]{Yangyang2023}%
  \BibitemOpen
  \bibfield  {author} {\bibinfo {author} {\bibfnamefont {Y.}~\bibnamefont
  {Ge}}, \bibinfo {author} {\bibfnamefont {X.}~\bibnamefont {Yu}}, \bibinfo
  {author} {\bibfnamefont {W.}~\bibnamefont {Xin}}, \bibinfo {author}
  {\bibfnamefont {Z.}~\bibnamefont {Wang}}, \bibinfo {author} {\bibfnamefont
  {Y.}~\bibnamefont {Zhang}}, \bibinfo {author} {\bibfnamefont
  {W.}~\bibnamefont {Zheng}}, \bibinfo {author} {\bibfnamefont
  {S.}~\bibnamefont {Li}}, \bibinfo {author} {\bibfnamefont {D.}~\bibnamefont
  {Lan}},\ and\ \bibinfo {author} {\bibfnamefont {Y.}~\bibnamefont {Yu}},\
  }\bibfield  {title} {\bibinfo {title} {{Efficient charging and discharging of
  a superconducting quantum battery through frequency-modulated driving}},\
  }\href {https://doi.org/10.1063/5.0161354} {\bibfield  {journal} {\bibinfo
  {journal} {Applied Physics Letters}\ }\textbf {\bibinfo {volume} {123}},\
  \bibinfo {pages} {154002} (\bibinfo {year} {2023})}\BibitemShut {NoStop}%
\bibitem [{\citenamefont {Mazzoncini}\ \emph {et~al.}(2023)\citenamefont
  {Mazzoncini}, \citenamefont {Cavina}, \citenamefont {Andolina}, \citenamefont
  {Erdman},\ and\ \citenamefont {Giovannetti}}]{Mazzoncini2023}%
  \BibitemOpen
  \bibfield  {author} {\bibinfo {author} {\bibfnamefont {F.}~\bibnamefont
  {Mazzoncini}}, \bibinfo {author} {\bibfnamefont {V.}~\bibnamefont {Cavina}},
  \bibinfo {author} {\bibfnamefont {G.~M.}\ \bibnamefont {Andolina}}, \bibinfo
  {author} {\bibfnamefont {P.~A.}\ \bibnamefont {Erdman}},\ and\ \bibinfo
  {author} {\bibfnamefont {V.}~\bibnamefont {Giovannetti}},\ }\bibfield
  {title} {\bibinfo {title} {Optimal control methods for quantum batteries},\
  }\href {https://doi.org/10.1103/PhysRevA.107.032218} {\bibfield  {journal}
  {\bibinfo  {journal} {Phys. Rev. A}\ }\textbf {\bibinfo {volume} {107}},\
  \bibinfo {pages} {032218} (\bibinfo {year} {2023})}\BibitemShut {NoStop}%
\bibitem [{\citenamefont {Beleño}\ \emph {et~al.}(2023)\citenamefont
  {Beleño}, \citenamefont {Santos},\ and\ \citenamefont
  {Barra}}]{micromaser2}%
  \BibitemOpen
  \bibfield  {author} {\bibinfo {author} {\bibfnamefont {Z.}~\bibnamefont
  {Beleño}}, \bibinfo {author} {\bibfnamefont {M.~F.}\ \bibnamefont
  {Santos}},\ and\ \bibinfo {author} {\bibfnamefont {F.}~\bibnamefont
  {Barra}},\ }\href@noop {} {\bibinfo {title} {Laser powered dissipative
  quantum batteries in atom-cavity qed}} (\bibinfo {year} {2023}),\ \Eprint
  {https://arxiv.org/abs/2310.09953} {arXiv:2310.09953 [quant-ph]} \BibitemShut
  {NoStop}%
\bibitem [{\citenamefont {Quach}\ and\ \citenamefont
  {Munro}(2020)}]{darkstates}%
  \BibitemOpen
  \bibfield  {author} {\bibinfo {author} {\bibfnamefont {J.~Q.}\ \bibnamefont
  {Quach}}\ and\ \bibinfo {author} {\bibfnamefont {W.~J.}\ \bibnamefont
  {Munro}},\ }\bibfield  {title} {\bibinfo {title} {Using dark states to charge
  and stabilize open quantum batteries},\ }\href
  {https://doi.org/10.1103/PhysRevApplied.14.024092} {\bibfield  {journal}
  {\bibinfo  {journal} {Phys. Rev. Appl.}\ }\textbf {\bibinfo {volume} {14}},\
  \bibinfo {pages} {024092} (\bibinfo {year} {2020})}\BibitemShut {NoStop}%
\bibitem [{\citenamefont {Feliu}\ and\ \citenamefont
  {Barra}(2024)}]{Feliu2024}%
  \BibitemOpen
  \bibfield  {author} {\bibinfo {author} {\bibfnamefont {D.}~\bibnamefont
  {Feliu}}\ and\ \bibinfo {author} {\bibfnamefont {F.}~\bibnamefont {Barra}},\
  }\bibfield  {title} {\bibinfo {title} {System-bath correlations and
  finite-time operation enhance the efficiency of a dissipative quantum
  battery},\ }\href
  {http://iopscience.iop.org/article/10.1088/2058-9565/ad4d1a} {\bibfield
  {journal} {\bibinfo  {journal} {Quantum Science and Technology}\ } (\bibinfo
  {year} {2024})},\ \Eprint {https://arxiv.org/abs/2310.09953}
  {arXiv:2310.09953 [quant-ph]} \BibitemShut {NoStop}%
\bibitem [{\citenamefont {Lenard}(1978{\natexlab{a}})}]{Gibbs}%
  \BibitemOpen
  \bibfield  {author} {\bibinfo {author} {\bibfnamefont {A.}~\bibnamefont
  {Lenard}},\ }\bibfield  {title} {\bibinfo {title} {Thermodynamical proof of
  the gibbs formula for elementary quantum systems},\ }\href
  {https://doi.org/10.1007/BF01011769} {\bibfield  {journal} {\bibinfo
  {journal} {Journal of Statistical Physics}\ }\textbf {\bibinfo {volume}
  {19}},\ \bibinfo {pages} {575} (\bibinfo {year}
  {1978}{\natexlab{a}})}\BibitemShut {NoStop}%
\bibitem [{\citenamefont {Allahverdyan}\ \emph {et~al.}(2004)\citenamefont
  {Allahverdyan}, \citenamefont {Balian},\ and\ \citenamefont
  {Nieuwenhuizen}}]{Ergotropy}%
  \BibitemOpen
  \bibfield  {author} {\bibinfo {author} {\bibfnamefont {A.~E.}\ \bibnamefont
  {Allahverdyan}}, \bibinfo {author} {\bibfnamefont {R.}~\bibnamefont
  {Balian}},\ and\ \bibinfo {author} {\bibfnamefont {T.~M.}\ \bibnamefont
  {Nieuwenhuizen}},\ }\bibfield  {title} {\bibinfo {title} {Maximal work
  extraction from finite quantum systems},\ }\href
  {https://doi.org/10.1209/epl/i2004-10101-2} {\bibfield  {journal} {\bibinfo
  {journal} {Europhysics Letters}\ }\textbf {\bibinfo {volume} {67}},\ \bibinfo
  {pages} {565} (\bibinfo {year} {2004})}\BibitemShut {NoStop}%
\bibitem [{\citenamefont {Piechocinska}(2000)}]{Piechocinska2000}%
  \BibitemOpen
  \bibfield  {author} {\bibinfo {author} {\bibfnamefont {B.}~\bibnamefont
  {Piechocinska}},\ }\bibfield  {title} {\bibinfo {title} {Information
  erasure},\ }\href {https://doi.org/10.1103/PhysRevA.61.062314} {\bibfield
  {journal} {\bibinfo  {journal} {Phys. Rev. A}\ }\textbf {\bibinfo {volume}
  {61}},\ \bibinfo {pages} {062314} (\bibinfo {year} {2000})}\BibitemShut
  {NoStop}%
\bibitem [{\citenamefont {Sagawa}\ and\ \citenamefont
  {Ueda}(2008)}]{Sagawa2008}%
  \BibitemOpen
  \bibfield  {author} {\bibinfo {author} {\bibfnamefont {T.}~\bibnamefont
  {Sagawa}}\ and\ \bibinfo {author} {\bibfnamefont {M.}~\bibnamefont {Ueda}},\
  }\bibfield  {title} {\bibinfo {title} {Second law of thermodynamics with
  discrete quantum feedback control},\ }\href
  {https://doi.org/10.1103/PhysRevLett.100.080403} {\bibfield  {journal}
  {\bibinfo  {journal} {Phys. Rev. Lett.}\ }\textbf {\bibinfo {volume} {100}},\
  \bibinfo {pages} {080403} (\bibinfo {year} {2008})}\BibitemShut {NoStop}%
\bibitem [{\citenamefont {Jacobs}(2009)}]{Jacobs}%
  \BibitemOpen
  \bibfield  {author} {\bibinfo {author} {\bibfnamefont {K.}~\bibnamefont
  {Jacobs}},\ }\bibfield  {title} {\bibinfo {title} {Second law of
  thermodynamics and quantum feedback control: Maxwell's demon with weak
  measurements},\ }\href {https://doi.org/10.1103/PhysRevA.80.012322}
  {\bibfield  {journal} {\bibinfo  {journal} {Phys. Rev. A}\ }\textbf {\bibinfo
  {volume} {80}},\ \bibinfo {pages} {012322} (\bibinfo {year}
  {2009})}\BibitemShut {NoStop}%
\bibitem [{\citenamefont {Funo}\ \emph {et~al.}(2013)\citenamefont {Funo},
  \citenamefont {Watanabe},\ and\ \citenamefont {Ueda}}]{Funo2013}%
  \BibitemOpen
  \bibfield  {author} {\bibinfo {author} {\bibfnamefont {K.}~\bibnamefont
  {Funo}}, \bibinfo {author} {\bibfnamefont {Y.}~\bibnamefont {Watanabe}},\
  and\ \bibinfo {author} {\bibfnamefont {M.}~\bibnamefont {Ueda}},\ }\bibfield
  {title} {\bibinfo {title} {Thermodynamic work gain from entanglement},\
  }\href {https://doi.org/10.1103/PhysRevA.88.052319} {\bibfield  {journal}
  {\bibinfo  {journal} {Phys. Rev. A}\ }\textbf {\bibinfo {volume} {88}},\
  \bibinfo {pages} {052319} (\bibinfo {year} {2013})}\BibitemShut {NoStop}%
\bibitem [{\citenamefont {Sagawa}\ and\ \citenamefont
  {Ueda}(2009)}]{Sagawa2009}%
  \BibitemOpen
  \bibfield  {author} {\bibinfo {author} {\bibfnamefont {T.}~\bibnamefont
  {Sagawa}}\ and\ \bibinfo {author} {\bibfnamefont {M.}~\bibnamefont {Ueda}},\
  }\bibfield  {title} {\bibinfo {title} {Minimal energy cost for thermodynamic
  information processing: Measurement and information erasure},\ }\href
  {https://doi.org/10.1103/PhysRevLett.102.250602} {\bibfield  {journal}
  {\bibinfo  {journal} {Phys. Rev. Lett.}\ }\textbf {\bibinfo {volume} {102}},\
  \bibinfo {pages} {250602} (\bibinfo {year} {2009})}\BibitemShut {NoStop}%
\bibitem [{\citenamefont {Jacobs}(2012)}]{Jacobs2}%
  \BibitemOpen
  \bibfield  {author} {\bibinfo {author} {\bibfnamefont {K.}~\bibnamefont
  {Jacobs}},\ }\bibfield  {title} {\bibinfo {title} {Quantum measurement and
  the first law of thermodynamics: The energy cost of measurement is the work
  value of the acquired information},\ }\href
  {https://doi.org/10.1103/PhysRevE.86.040106} {\bibfield  {journal} {\bibinfo
  {journal} {Phys. Rev. E}\ }\textbf {\bibinfo {volume} {86}},\ \bibinfo
  {pages} {040106} (\bibinfo {year} {2012})}\BibitemShut {NoStop}%
\bibitem [{\citenamefont {Francisca}\ \emph {et~al.}(2017)\citenamefont
  {Francisca}, \citenamefont {Goold}, \citenamefont {Plastina},\ and\
  \citenamefont {Paternostro}}]{Daemonic}%
  \BibitemOpen
  \bibfield  {author} {\bibinfo {author} {\bibfnamefont {G.}~\bibnamefont
  {Francisca}}, \bibinfo {author} {\bibfnamefont {J.}~\bibnamefont {Goold}},
  \bibinfo {author} {\bibfnamefont {F.}~\bibnamefont {Plastina}},\ and\
  \bibinfo {author} {\bibfnamefont {M.}~\bibnamefont {Paternostro}},\
  }\bibfield  {title} {\bibinfo {title} {Daemonic ergotropy: enhanced work
  extraction from quantum correlations},\ }\href
  {https://doi.org/10.1038/s41534-017-0012-8} {\bibfield  {journal} {\bibinfo
  {journal} {npj Quantum Information}\ }\textbf {\bibinfo {volume} {3}},\
  \bibinfo {pages} {12} (\bibinfo {year} {2017})}\BibitemShut {NoStop}%
\bibitem [{\citenamefont {Elouard}\ \emph {et~al.}(2017)\citenamefont
  {Elouard}, \citenamefont {Herrera-Mart\'{\i}}, \citenamefont {Huard},\ and\
  \citenamefont {Auff\`eves}}]{Auffeves2017}%
  \BibitemOpen
  \bibfield  {author} {\bibinfo {author} {\bibfnamefont {C.}~\bibnamefont
  {Elouard}}, \bibinfo {author} {\bibfnamefont {D.}~\bibnamefont
  {Herrera-Mart\'{\i}}}, \bibinfo {author} {\bibfnamefont {B.}~\bibnamefont
  {Huard}},\ and\ \bibinfo {author} {\bibfnamefont {A.}~\bibnamefont
  {Auff\`eves}},\ }\bibfield  {title} {\bibinfo {title} {Extracting work from
  quantum measurement in maxwell's demon engines},\ }\href
  {https://doi.org/10.1103/PhysRevLett.118.260603} {\bibfield  {journal}
  {\bibinfo  {journal} {Phys. Rev. Lett.}\ }\textbf {\bibinfo {volume} {118}},\
  \bibinfo {pages} {260603} (\bibinfo {year} {2017})}\BibitemShut {NoStop}%
\bibitem [{\citenamefont {Manzano}\ \emph {et~al.}(2018)\citenamefont
  {Manzano}, \citenamefont {Plastina},\ and\ \citenamefont
  {Zambrini}}]{Manzano2018}%
  \BibitemOpen
  \bibfield  {author} {\bibinfo {author} {\bibfnamefont {G.}~\bibnamefont
  {Manzano}}, \bibinfo {author} {\bibfnamefont {F.}~\bibnamefont {Plastina}},\
  and\ \bibinfo {author} {\bibfnamefont {R.}~\bibnamefont {Zambrini}},\
  }\bibfield  {title} {\bibinfo {title} {Optimal work extraction and
  thermodynamics of quantum measurements and correlations},\ }\href
  {https://doi.org/10.1103/PhysRevLett.121.120602} {\bibfield  {journal}
  {\bibinfo  {journal} {Phys. Rev. Lett.}\ }\textbf {\bibinfo {volume} {121}},\
  \bibinfo {pages} {120602} (\bibinfo {year} {2018})}\BibitemShut {NoStop}%
\bibitem [{\citenamefont {Bernards}\ \emph {et~al.}(2019)\citenamefont
  {Bernards}, \citenamefont {Kleinmann}, \citenamefont {Gühne},\ and\
  \citenamefont {Paternostro}}]{POVM}%
  \BibitemOpen
  \bibfield  {author} {\bibinfo {author} {\bibfnamefont {F.}~\bibnamefont
  {Bernards}}, \bibinfo {author} {\bibfnamefont {M.}~\bibnamefont {Kleinmann}},
  \bibinfo {author} {\bibfnamefont {O.}~\bibnamefont {Gühne}},\ and\ \bibinfo
  {author} {\bibfnamefont {M.}~\bibnamefont {Paternostro}},\ }\bibfield
  {title} {\bibinfo {title} {Daemonic ergotropy: Generalised measurements and
  multipartite settings},\ }\bibfield  {journal} {\bibinfo  {journal}
  {Entropy}\ }\textbf {\bibinfo {volume} {21}},\ \href
  {https://doi.org/10.3390/e21080771} {10.3390/e21080771} (\bibinfo {year}
  {2019})\BibitemShut {NoStop}%
\bibitem [{\citenamefont {Landauer}(1961)}]{Landauer}%
  \BibitemOpen
  \bibfield  {author} {\bibinfo {author} {\bibfnamefont {R.}~\bibnamefont
  {Landauer}},\ }\bibfield  {title} {\bibinfo {title} {Irreversibility and heat
  generation in the computing process},\ }\href
  {https://doi.org/10.1147/rd.53.0183} {\bibfield  {journal} {\bibinfo
  {journal} {IBM Journal of Research and Development}\ }\textbf {\bibinfo
  {volume} {5}},\ \bibinfo {pages} {183} (\bibinfo {year} {1961})}\BibitemShut
  {NoStop}%
\bibitem [{\citenamefont {Plenio}\ and\ \citenamefont
  {Vitelli}(2001)}]{Landauerreview}%
  \BibitemOpen
  \bibfield  {author} {\bibinfo {author} {\bibfnamefont {M.~B.}\ \bibnamefont
  {Plenio}}\ and\ \bibinfo {author} {\bibfnamefont {V.}~\bibnamefont
  {Vitelli}},\ }\bibfield  {title} {\bibinfo {title} {The physics of
  forgetting: Landauer's erasure principle and information theory},\ }\href
  {https://doi.org/10.1080/00107510010018916} {\bibfield  {journal} {\bibinfo
  {journal} {Contemporary Physics}\ }\textbf {\bibinfo {volume} {42}},\
  \bibinfo {pages} {25} (\bibinfo {year} {2001})}\BibitemShut {NoStop}%
\bibitem [{\citenamefont {Barkeshli}(2005)}]{Barkeshli2005}%
  \BibitemOpen
  \bibfield  {author} {\bibinfo {author} {\bibfnamefont {M.}~\bibnamefont
  {Barkeshli}},\ }\bibfield  {title} {\bibinfo {title} {Dissipationless
  information erasure and landauer's principle},\ }\href
  {https://api.semanticscholar.org/CorpusID:118919743} {\bibfield  {journal}
  {\bibinfo  {journal} {arXiv: Statistical Mechanics}\ } (\bibinfo {year}
  {2005})}\BibitemShut {NoStop}%
\bibitem [{\citenamefont {Bérut}\ \emph {et~al.}(2012)\citenamefont {Bérut},
  \citenamefont {Arakelyan}, \citenamefont {Petrosyan}, \citenamefont
  {Ciliberto}, \citenamefont {Dillenschneider},\ and\ \citenamefont
  {Lutz}}]{LandauerExperiment}%
  \BibitemOpen
  \bibfield  {author} {\bibinfo {author} {\bibfnamefont {A.}~\bibnamefont
  {Bérut}}, \bibinfo {author} {\bibfnamefont {A.}~\bibnamefont {Arakelyan}},
  \bibinfo {author} {\bibfnamefont {A.}~\bibnamefont {Petrosyan}}, \bibinfo
  {author} {\bibfnamefont {S.}~\bibnamefont {Ciliberto}}, \bibinfo {author}
  {\bibfnamefont {R.}~\bibnamefont {Dillenschneider}},\ and\ \bibinfo {author}
  {\bibfnamefont {E.}~\bibnamefont {Lutz}},\ }\bibfield  {title} {\bibinfo
  {title} {Experimental verification of landauer’s principle linking
  information and thermodynamics},\ }\href
  {https://doi.org/https://doi.org/10.1038/nature10872} {\bibfield  {journal}
  {\bibinfo  {journal} {Nature}\ }\textbf {\bibinfo {volume} {483}},\ \bibinfo
  {pages} {187} (\bibinfo {year} {2012})}\BibitemShut {NoStop}%
\bibitem [{\citenamefont {Maruyama}\ \emph {et~al.}(2009)\citenamefont
  {Maruyama}, \citenamefont {Nori},\ and\ \citenamefont
  {Vedral}}]{Maruyama2009}%
  \BibitemOpen
  \bibfield  {author} {\bibinfo {author} {\bibfnamefont {K.}~\bibnamefont
  {Maruyama}}, \bibinfo {author} {\bibfnamefont {F.}~\bibnamefont {Nori}},\
  and\ \bibinfo {author} {\bibfnamefont {V.}~\bibnamefont {Vedral}},\
  }\bibfield  {title} {\bibinfo {title} {Colloquium: The physics of maxwell's
  demon and information},\ }\href {https://doi.org/10.1103/RevModPhys.81.1}
  {\bibfield  {journal} {\bibinfo  {journal} {Rev. Mod. Phys.}\ }\textbf
  {\bibinfo {volume} {81}},\ \bibinfo {pages} {1} (\bibinfo {year}
  {2009})}\BibitemShut {NoStop}%
\bibitem [{\citenamefont {Parrondo}\ \emph {et~al.}(2015)\citenamefont
  {Parrondo}, \citenamefont {Horowitz},\ and\ \citenamefont {Sagawa}}]{ToIrev}%
  \BibitemOpen
  \bibfield  {author} {\bibinfo {author} {\bibfnamefont {J.~M.~R.}\
  \bibnamefont {Parrondo}}, \bibinfo {author} {\bibfnamefont {J.~M.}\
  \bibnamefont {Horowitz}},\ and\ \bibinfo {author} {\bibfnamefont
  {T.}~\bibnamefont {Sagawa}},\ }\bibfield  {title} {\bibinfo {title}
  {Thermodynamics of information},\ }\href
  {https://doi.org/https://doi.org/10.1038/nphys3230} {\bibfield  {journal}
  {\bibinfo  {journal} {Nature Physics}\ }\textbf {\bibinfo {volume} {11}},\
  \bibinfo {pages} {131} (\bibinfo {year} {2015})}\BibitemShut {NoStop}%
\bibitem [{\citenamefont {Perarnau-Llobet}\ \emph {et~al.}(2015)\citenamefont
  {Perarnau-Llobet}, \citenamefont {Hovhannisyan}, \citenamefont {Huber},
  \citenamefont {Skrzypczyk}, \citenamefont {Brunner},\ and\ \citenamefont
  {Ac\'{\i}n}}]{Perarnau-Llobet2015}%
  \BibitemOpen
  \bibfield  {author} {\bibinfo {author} {\bibfnamefont {M.}~\bibnamefont
  {Perarnau-Llobet}}, \bibinfo {author} {\bibfnamefont {K.~V.}\ \bibnamefont
  {Hovhannisyan}}, \bibinfo {author} {\bibfnamefont {M.}~\bibnamefont {Huber}},
  \bibinfo {author} {\bibfnamefont {P.}~\bibnamefont {Skrzypczyk}}, \bibinfo
  {author} {\bibfnamefont {N.}~\bibnamefont {Brunner}},\ and\ \bibinfo {author}
  {\bibfnamefont {A.}~\bibnamefont {Ac\'{\i}n}},\ }\bibfield  {title} {\bibinfo
  {title} {Extractable work from correlations},\ }\href
  {https://doi.org/10.1103/PhysRevX.5.041011} {\bibfield  {journal} {\bibinfo
  {journal} {Phys. Rev. X}\ }\textbf {\bibinfo {volume} {5}},\ \bibinfo {pages}
  {041011} (\bibinfo {year} {2015})}\BibitemShut {NoStop}%
\bibitem [{\citenamefont {Bresque}\ \emph {et~al.}(2021)\citenamefont
  {Bresque}, \citenamefont {Camati}, \citenamefont {Rogers}, \citenamefont
  {Murch}, \citenamefont {Jordan},\ and\ \citenamefont {Auff\`eves}}]{Fuel}%
  \BibitemOpen
  \bibfield  {author} {\bibinfo {author} {\bibfnamefont {L.}~\bibnamefont
  {Bresque}}, \bibinfo {author} {\bibfnamefont {P.~A.}\ \bibnamefont {Camati}},
  \bibinfo {author} {\bibfnamefont {S.}~\bibnamefont {Rogers}}, \bibinfo
  {author} {\bibfnamefont {K.}~\bibnamefont {Murch}}, \bibinfo {author}
  {\bibfnamefont {A.~N.}\ \bibnamefont {Jordan}},\ and\ \bibinfo {author}
  {\bibfnamefont {A.}~\bibnamefont {Auff\`eves}},\ }\bibfield  {title}
  {\bibinfo {title} {Two-qubit engine fueled by entanglement and local
  measurements},\ }\href {https://doi.org/10.1103/PhysRevLett.126.120605}
  {\bibfield  {journal} {\bibinfo  {journal} {Phys. Rev. Lett.}\ }\textbf
  {\bibinfo {volume} {126}},\ \bibinfo {pages} {120605} (\bibinfo {year}
  {2021})}\BibitemShut {NoStop}%
\bibitem [{Note1()}]{Note1}%
  \BibitemOpen
  \bibinfo {note} {A smooth disconnecting protocol was studied in~\cite
  {Feliu2024}.}\BibitemShut {Stop}%
\bibitem [{\citenamefont {Lenard}(1978{\natexlab{b}})}]{Lenard}%
  \BibitemOpen
  \bibfield  {author} {\bibinfo {author} {\bibfnamefont {A.}~\bibnamefont
  {Lenard}},\ }\bibfield  {title} {\bibinfo {title} {Thermodynamical proof of
  the gibbs formula for elementary quantum systems},\ }\href
  {https://doi.org/10.1007/BF01011769} {\bibfield  {journal} {\bibinfo
  {journal} {Journal of Statistical Physics}\ }\textbf {\bibinfo {volume}
  {19}},\ \bibinfo {pages} {575} (\bibinfo {year}
  {1978}{\natexlab{b}})}\BibitemShut {NoStop}%
\bibitem [{\citenamefont {Cottet}\ \emph {et~al.}(2017)\citenamefont {Cottet},
  \citenamefont {Jezouin}, \citenamefont {Bretheau}, \citenamefont
  {Campagne-Ibarcq}, \citenamefont {Ficheux}, \citenamefont {Anders},
  \citenamefont {Auffèves}, \citenamefont {Azouit}, \citenamefont {Rouchon},\
  and\ \citenamefont {Huard}}]{Cottet2017}%
  \BibitemOpen
  \bibfield  {author} {\bibinfo {author} {\bibfnamefont {N.}~\bibnamefont
  {Cottet}}, \bibinfo {author} {\bibfnamefont {S.}~\bibnamefont {Jezouin}},
  \bibinfo {author} {\bibfnamefont {L.}~\bibnamefont {Bretheau}}, \bibinfo
  {author} {\bibfnamefont {P.}~\bibnamefont {Campagne-Ibarcq}}, \bibinfo
  {author} {\bibfnamefont {Q.}~\bibnamefont {Ficheux}}, \bibinfo {author}
  {\bibfnamefont {J.}~\bibnamefont {Anders}}, \bibinfo {author} {\bibfnamefont
  {A.}~\bibnamefont {Auffèves}}, \bibinfo {author} {\bibfnamefont
  {R.}~\bibnamefont {Azouit}}, \bibinfo {author} {\bibfnamefont
  {P.}~\bibnamefont {Rouchon}},\ and\ \bibinfo {author} {\bibfnamefont
  {B.}~\bibnamefont {Huard}},\ }\bibfield  {title} {\bibinfo {title} {Observing
  a quantum maxwell demon at work},\ }\href
  {https://doi.org/10.1073/pnas.1704827114} {\bibfield  {journal} {\bibinfo
  {journal} {Proceedings of the National Academy of Sciences}\ }\textbf
  {\bibinfo {volume} {114}},\ \bibinfo {pages} {7561} (\bibinfo {year}
  {2017})}\BibitemShut {NoStop}%
\bibitem [{\citenamefont {Mitchison}\ \emph {et~al.}(2021)\citenamefont
  {Mitchison}, \citenamefont {Goold}, ,\ and\ \citenamefont
  {Prior}}]{ChargwithLFb}%
  \BibitemOpen
  \bibfield  {author} {\bibinfo {author} {\bibfnamefont {M.}~\bibnamefont
  {Mitchison}}, \bibinfo {author} {\bibfnamefont {J.}~\bibnamefont {Goold}}, ,\
  and\ \bibinfo {author} {\bibfnamefont {J.}~\bibnamefont {Prior}},\ }\bibfield
   {title} {\bibinfo {title} {Charging a quantum battery with linear feedback
  control},\ }\href {https://doi.org/10.1038/s41534-017-0012-8} {\bibfield
  {journal} {\bibinfo  {journal} {Quantum}\ }\textbf {\bibinfo {volume} {5}},\
  \bibinfo {pages} {500} (\bibinfo {year} {2021})}\BibitemShut {NoStop}%
\bibitem [{\citenamefont {Morrone}\ \emph
  {et~al.}(2023{\natexlab{b}})\citenamefont {Morrone}, \citenamefont {Rossi},\
  and\ \citenamefont {Genoni}}]{DaemonicCont}%
  \BibitemOpen
  \bibfield  {author} {\bibinfo {author} {\bibfnamefont {D.}~\bibnamefont
  {Morrone}}, \bibinfo {author} {\bibfnamefont {M.~A.}\ \bibnamefont {Rossi}},\
  and\ \bibinfo {author} {\bibfnamefont {M.~G.}\ \bibnamefont {Genoni}},\
  }\bibfield  {title} {\bibinfo {title} {Daemonic ergotropy in continuously
  monitored open quantum batteries},\ }\href
  {https://doi.org/10.1103/PhysRevApplied.20.044073} {\bibfield  {journal}
  {\bibinfo  {journal} {Phys. Rev. Appl.}\ }\textbf {\bibinfo {volume} {20}},\
  \bibinfo {pages} {044073} (\bibinfo {year} {2023}{\natexlab{b}})}\BibitemShut
  {NoStop}%
\bibitem [{\citenamefont {Yan}\ and\ \citenamefont {Jing}(2023)}]{Yan2023}%
  \BibitemOpen
  \bibfield  {author} {\bibinfo {author} {\bibfnamefont {J.-s.}\ \bibnamefont
  {Yan}}\ and\ \bibinfo {author} {\bibfnamefont {J.}~\bibnamefont {Jing}},\
  }\bibfield  {title} {\bibinfo {title} {Charging by quantum measurement},\
  }\href {https://doi.org/10.1103/PhysRevApplied.19.064069} {\bibfield
  {journal} {\bibinfo  {journal} {Phys. Rev. Appl.}\ }\textbf {\bibinfo
  {volume} {19}},\ \bibinfo {pages} {064069} (\bibinfo {year}
  {2023})}\BibitemShut {NoStop}%
\bibitem [{\citenamefont {Esposito}\ and\ \citenamefont {den
  Broeck}(2011)}]{Esposito2011}%
  \BibitemOpen
  \bibfield  {author} {\bibinfo {author} {\bibfnamefont {M.}~\bibnamefont
  {Esposito}}\ and\ \bibinfo {author} {\bibfnamefont {C.~V.}\ \bibnamefont {den
  Broeck}},\ }\bibfield  {title} {\bibinfo {title} {Second law and landauer
  principle far from equilibrium},\ }\href
  {https://doi.org/10.1209/0295-5075/95/40004} {\bibfield  {journal} {\bibinfo
  {journal} {Europhysics Letters}\ }\textbf {\bibinfo {volume} {95}},\ \bibinfo
  {pages} {40004} (\bibinfo {year} {2011})}\BibitemShut {NoStop}%
\bibitem [{\citenamefont {Groenewold}(1971)}]{Groenewold1971}%
  \BibitemOpen
  \bibfield  {author} {\bibinfo {author} {\bibfnamefont {H.~J.}\ \bibnamefont
  {Groenewold}},\ }\bibfield  {title} {\bibinfo {title} {A problem of
  information gain by quantal measurements},\ }\href
  {https://doi.org/10.1007/BF00815357} {\bibfield  {journal} {\bibinfo
  {journal} {Int. J. Theor. Phys.}\ }\textbf {\bibinfo {volume} {4}},\ \bibinfo
  {pages} {327} (\bibinfo {year} {1971})}\BibitemShut {NoStop}%
\bibitem [{\citenamefont {Nielsen}\ and\ \citenamefont {Chuang}(2000)}]{QCQI}%
  \BibitemOpen
  \bibfield  {author} {\bibinfo {author} {\bibfnamefont {M.~A.}\ \bibnamefont
  {Nielsen}}\ and\ \bibinfo {author} {\bibfnamefont {I.~L.}\ \bibnamefont
  {Chuang}},\ }\href@noop {} {\emph {\bibinfo {title} {Quantum Computation and
  Quantum Information}}}\ (\bibinfo  {publisher} {Cambridge University Press},\
  \bibinfo {year} {2000})\BibitemShut {NoStop}%
\bibitem [{\citenamefont {Feynman}(1986)}]{CNOT}%
  \BibitemOpen
  \bibfield  {author} {\bibinfo {author} {\bibfnamefont {R.}~\bibnamefont
  {Feynman}},\ }\bibfield  {title} {\bibinfo {title} {Quantum mechanical
  computers},\ }\href {https://doi.org/10.1007/BF01886518} {\bibfield
  {journal} {\bibinfo  {journal} {Foundations of Physics}\ }\textbf {\bibinfo
  {volume} {16}},\ \bibinfo {pages} {507} (\bibinfo {year} {1986})}\BibitemShut
  {NoStop}%
\end{thebibliography}%

\end{document}